\PassOptionsToPackage{square,comma,numbers,sort&compress}{natbib}  

 \documentclass[final,3p,times]{elsarticle}

\usepackage{lineno}
\usepackage{smartdiagram}  
\usepackage{cancel}
\usepackage{multicol}
\usepackage{booktabs} 
\usepackage{etoolbox}  
\usepackage{listings} 

\usepackage{framed} 
 
\usepackage{nomencl} 
\makenomenclature
\renewcommand*\nompreamble{\begin{multicols}{2}}
\renewcommand*\nompostamble{\end{multicols}}

\renewcommand\nomgroup[1]{%
  \item[\bfseries
  \ifstrequal{#1}{A}{ Abbrevations }{%
  \ifstrequal{#1}{D}{ Dimensionless Variables}{%
  \ifstrequal{#1}{S}{ Symbols}{%
  \ifstrequal{#1}{B}{ Subscipts}{%
  \ifstrequal{#1}{P}{ Superscipts}}}}}%
]}

\makeindex

\usepackage[T1]{fontenc}

\usepackage[utf8]{inputenc}

\usepackage[english]{babel}
\addto\captionsenglish{} 
\usepackage{amsmath, amsthm, amsfonts, amssymb}
\usepackage{bbm}
\usepackage{csquotes}

\usepackage[tableposition=top]{caption}

\usepackage{graphicx}
\usepackage{subcaption} 

\usepackage{setspace}
\usepackage{verbatim}

\usepackage{lscape}

\usepackage{graphicx}
\usepackage{amssymb}

\usepackage{tikz}
\usetikzlibrary{shapes,arrows}
\tikzstyle{arrow} = [thick,->,>=stealth]

\usepackage{natbib} 

\usepackage{lineno} 

\newcommand{\chapternote}[1]{{%
  \let\thempfn\relax
   \footnotetext[0]{#1}
}}

\usepackage{footnote} 

\usepackage{hyperref} 
\hypersetup{urlcolor=blue, 
    citecolor=blue,
    filecolor=blue,
    linkcolor=blue,
    urlcolor=blue,
colorlinks=true} 

\newcommand{\bs}[1]{\boldsymbol{#1}}

\newcommand{\pr}[1]{\frac{\partial}{\partial #1}}

\newcommand{\eq}{\text{eq}}

\def\etal.{et\penalty50\ al.}

\newcommand{\h}{h}
\newcommand{\lbmh}{\tilde{\h}}

\newcommand{\thei}{\mathrm{i}}
\newcommand{\bY}{{\bs\Upsilon}}

\usepackage{todonotes}
\makeatletter
\renewcommand{\@todonotes@drawLineToRightMargin}{%
\if@todonotes@line%
	\begin{tikzpicture}[remember picture, overlay]%
		\draw[connectstyle]%
		(inText)%
		-- ([yshift=-0.2cm + \@todonotes@tickmarkheight] inText)%
		-- ([xshift=-0.4cm, yshift=-0.2cm + \@todonotes@tickmarkheight] inNote.west |- inText)%
		-- (inNote.west);%
	\end{tikzpicture}%
\fi}
\makeatother

\usepackage{acro}
    
\DeclareAcronym{LBM}{
    short = LBM,
    long  = lattice Boltzmann method,
    class = nomencl
    }
    
\DeclareAcronym{FEM}{
    short = FEM,
    long  = finite element method,
    class = nomencl
    }  
    
\DeclareAcronym{FD}{
    short = FD,
    long  = finite difference,
    class = nomencl
    }         

\DeclareAcronym{IBM}{
    short = IBM,
    long  = immersed boundary method,
    class = nomencl
    }

\DeclareAcronym{NS}{
    short = NS,
    long  = Navier Stokes,
    class = nomencl
    }
    
\DeclareAcronym{ADE}{
    short = ADE,
    long  = advection-diffusion equation,
    class = nomencl
    } 
    
\DeclareAcronym{ADRE}{
    short = ADRE,
    long  = advection-diffusion-reaction equation,
    class = nomencl
    } 
            
\DeclareAcronym{CM}{
    short = CM,
    long  = central moments,
    class = nomencl
    }    

\DeclareAcronym{SRT}{
    short = SRT,
    long  = single relaxation time,
    class = nomencl
    } 

\DeclareAcronym{MRT}{
    short = MRT,
    long  = multiple relaxation time,
    class = nomencl
    } 
    
\DeclareAcronym{TRT}{
    short = TRT,
    long  = two relaxation time,
    class = nomencl
    } 

\DeclareAcronym{DBE}{
    short = DBE,
    long  = discrete Boltzmann equation,
    class = nomencl
    }  
    
\DeclareAcronym{PDE}{
    short = PDE,
    long  = partial differential equation,
    class = nomencl
    }  

\DeclareAcronym{ODE}{
    short = ODE,
    long  = ordinary differential equation,
    class = nomencl
    }  
                   
\DeclareAcronym{Pr}{
    short = $Pr$,
    long  = Prandtl number,
    class = nomencl
    }

\DeclareAcronym{Da}{
    short = $\mathbf{Da}$,
    long  = Damk{\"o}hler number,
    class = nomencl
    }               

\DeclareAcronym{Fo}{
    short = $\mathbf{Fo}$,
    long  = Fourier number,
    class = nomencl
    }     
    \DeclareAcronym{Pe}{
    short = $\mathbf{Pe}$,
    long  = P{\'e}clet number,
    class = nomencl
    }

\nomenclature[A]{\acs{LBM}}{\acl{LBM}}
\nomenclature[A]{\acs{ADE}}{\acl{ADE}}
\nomenclature[A]{\acs{ADRE}}{\acl{ADRE}}

\nomenclature[A]{\acs{SRT}}{\acl{SRT}}
\nomenclature[A]{\acs{TRT}}{\acl{TRT}}
\nomenclature[A]{\acs{MRT}}{\acl{MRT}}

\nomenclature[A]{\acs{DBE}}{\acl{DBE}}

\nomenclature[A]{\acs{PDE}}{\acl{PDE}}
\nomenclature[A]{\acs{ODE}}{\acl{ODE}}

\nomenclature[A]{\acs{FEM} }{\acl{FEM} }
\nomenclature[A]{\acs{FD} }{\acl{FD} }

\nomenclature[S]{$\h_i$}{$i$-th distribution function for the $\phi$ field}

\nomenclature[S]{$\Upsilon_{i}$}{$i$-th raw moment of the distribution function}
\nomenclature[S]{$M$}{Macroscopic diffusion coefficient}

\nomenclature[S]{$\mathbb{M}$}{Transformation matrix}
\nomenclature[S]{$\mathbb{S}$}{Relaxation matrix}

\nomenclature[S]{$\mathbf{x}$}{Position vector}
\nomenclature[S]{$\mathbf{e}$}{Characteristic lattice velocity}
\nomenclature[S]{$\mathbf{u}$}{Macroscopic velocity vector}
\nomenclature[S]{$Q$}{Macroscopic source term}
\nomenclature[S]{$\phi$}{Density of the scalar field}

\nomenclature[S]{$c_s$}{Lattice speed of sound}
\nomenclature[S]{$\thei$}{Imaginary number}
\nomenclature[S]{$\tau$}{Relaxation time}
\nomenclature[S]{$\omega_i$}{$i$-th relaxation frequency}

\nomenclature[S]{$T$}{Number of time iterations (time steps)}
\nomenclature[S]{$L$}{Number of lattice elements per length}

\nomenclature[S]{$\Delta x$}{Spatial resolution ($1/L$)}
\nomenclature[S]{$\Delta t$}{Temporal resolution ($1/T$)}

\nomenclature[P]{$eq$}{Equilibrium}
\nomenclature[P]{$\wedge$}{Variables from the next time step (t+1)}
\nomenclature[P]{$\star$}{Post-collision variable}
\nomenclature[P]{$\backsim$}{redefined (shifted) distribution function or its moment}

\nomenclature[D]{\acs{Fo}}{\acl{Fo}}
\nomenclature[D]{\acs{Da}}{\acl{Da}}
\nomenclature[D]{\acs{Pe}}{\acl{Pe}}

\usepackage[capitalise, noabbrev]{cleveref}

\usepackage{siunitx} 
\usepackage{nicefrac} 

\journal{Elsevier}
\begin{document}

\begin{frontmatter}

\title{Revisiting the second-order convergence of \\ the lattice Boltzmann method with reaction-type source terms}




\author[First,Second]{Grzegorz Gruszczy\'nski\corref{cor1}
}
\ead{ggruszczynski@gmail.com}

\author[Second]{Micha\l{} Dzikowski
}
\ead{mjdzikowski@gmail.com}

\author[Third]{\L{}ukasz \L{}aniewski-Wo\l{}\l{}k }
\cortext[cor1]{Corresponding author.}

\address[First]{Institute of Aeronautics and Applied Mechanics, Warsaw University of Technology, Warszawa, Poland}
\address[Second]{Interdisciplinary Centre for Mathematical and Computational Modelling, University of Warsaw, Warszawa, Poland}
\address[Third]{School of Mechanical and Mining Engineering, The University of Queensland, St Lucia, Australia}

\begin{abstract}
This study analyses an approach to consistently recover the second-order convergence of the \ac{LBM}, 
which is frequently degraded by an improper discretisation of the required source terms.
The current work focuses on advection-diffusion models, in which the source terms are dependent on the intensity of transported fields.
Such terms can be observed in reaction-type equations used in heat and mass transfer problems or multiphase flows.
The investigated scheme is applicable to a wide range of formulations within the \ac{LBM} framework.
All considered source terms are interpreted as contributions to the zeroth-moment of the distribution function.
These account for sources in a scalar field, such as density, concentration, temperature or a phase field.
Further application of this work can be found in the method of manufactured solutions or in the immersed boundary method.

This paper is dedicated to three aspects regarding proper inclusion of the source term in \ac{LBM} schemes.
Firstly, it identifies the differences observed between the ways in which source terms are included in the \ac{LBM} schemes present in the literature.
The algebraic manipulations are explicitly presented in this paper to clarify the observed differences, and to identify their origin.
Secondly, it analyses in full detail, the implicit relation between the value of the transported macroscopic field, and the sum of the \ac{LBM} densities.
This relation is valid for any source term discretization scheme.
It is a crucial ingredient for preserving the second-order convergence in the case of complex source terms.
Moreover, three equivalent forms of the second-order accurate collision operator are presented.
Finally, closed form solutions of this implicit relation are shown for a variety of  common models, including general linear and second order terms; population growth models, such as the Logistic or Gompertz model and the Allen-Cahn equation.

The second-order convergence of the proposed \ac{LBM} schemes is verified on both linear and non-linear source terms.
The pitfalls of the commonly used acoustic and diffusive scalings are identified and discussed.
Furthermore, for a simplified case, the competing errors are shown visually with isolines of error in the space of spatial and temporal resolutions.
\end{abstract}


\end{frontmatter}


\acresetall 
\section{Introduction}\label{sec:Intro}
The \ac{LBM} is a widely used numerical scheme for solving both the \ac{NS} equations and \ac{ADRE}.
Its popularity has significantly risen in the recent three decades due to its ability to handle complex boundary shapes and its relative ease of implementation and parallelisation.
In general, the explicitness of the scheme and the second-order convergence of the \ac{LBM} is known the literature~\cite{Doolen1998,Kruger2017}.
Unfortunately, it may be easily degraded through inconsistent discretisation of the source terms. 
This may lead to excessive computational requirements to achieve accuracies observed in other formulations.

To simplify derivations, the scope of this paper is narrowed to an advection-diffusion-reaction equation,
\begin{linenomath}\begin{align}
\pr t\phi+\nabla\cdot(\boldsymbol{u}\phi)=\nabla\cdot(M\nabla\phi)+Q(\phi, \bs x,t),\label{eq:ADE}
\end{align}\end{linenomath}
where $\phi$ denotes the scalar field, $\bs u$ refers to an advection velocity, $M$ is a diffusion coefficient, and $Q$ is the source term.
In this work, a consistent discretisation of the source term is discussed in order to recover the second-order convergence of the underlying scheme.
The discussion is focused on the situation in which $Q$ is dependent on $\phi$ itself, i.e., $Q=Q(\phi)$.
To the best of the authors' knowledge, such a form has not been analysed in the literature.
Integration of the \ac{DBE} with trapezoidal rule leads to implicit expressions. 
To transform them into an explicit scheme, a shift (redefinition) of variables is used~\cite{Doolen1998,Kruger2017}.
The resulting relation between the macroscopic field, $\phi$, and the shifted one, $\tilde\phi$, can be expressed as,
\begin{linenomath}\begin{equation}
\phi - \dfrac{1}{2}Q(\phi,\bs x,t) = \tilde\phi.
\label{eq:intro_phi_from_phi}
\end{equation}\end{linenomath}
However, in the case of $Q$ dependent on the $\phi$, the~\cref{eq:intro_phi_from_phi} is implicit itself and potentially non-linear. 
The detailed analysis of this issue is the primary concern of the current work.
We highlight the often neglected details of the derivation of the \ac{LBM} in the aforementioned scenario.
Next, we exemplify the proper treatment of the implicit, 
scalar source term in \ac{ADRE} with an appropriate redefinition of variables and initialisation of the \ac{LBM} densities.

Finally, we clarify the potential confusion related to the way in which the convergence \ac{LBM} is investigated.
In classical numerical methods, spatial convergence is tested by varying the domain resolution while holding a fine, fixed time step.
The idea is to keep the error of temporal discretisation much smaller than the spatial one or vice versa.
Due to the relatively narrow range of parameters for which \ac{LBM} is stable, its convergence is usually investigated for a special relationship between spatial and temporal scales.
Usually, these relationships follow the rules of so-called acoustic and diffusive scaling.
As a consequence, a biased view of convergence may be reported by looking solely at either spatial or temporal resolution, while the corresponding errors can converge at different rates.
The naive discretisation of the source term (assuming $\phi=\tilde\phi$) reduces the order of convergence in the case of acoustic scaling. 
On the other hand, we will demonstrate that the same order of convergence can still be observed even for the improper treatment of the source term under diffusive scaling. 

\subsection{Discussion on the state of the art}\label{sec:state_of_the_art}
From an application point-of-view, there is a wide range of physical phenomena which can be described by a form of the advection-diffusion-reaction equation.
Models based on the \ac{LBM} were developed for processes such as
multicomponent reaction \cite{YuHang2017}, 
combustion \cite{Lin2016}, 
solid and fluid dissolution related to underground $CO_2$ storage \cite{Aursjo2015,An2021}, 
crystallization and melting \cite{Huang2013,Huang2015a}
and heat transfer \cite{Doolen1998,Guo2007,Karlin2013,Fei2018a,Fei2018c,Premnath2018cascadedSources}.
Last but not least, the  numerical solvers can be verified by the method of manufactured solutions \cite{salari2000code}
in which the source term modelling is extensively used.
Selected studies involving the use of \ac{LBM} are listed below to present both the scope of modelling approaches, and applicability of the concepts discussed in this work. 

One of the earliest studies in which the \ac{LBM} framework was applied to simulate reaction-diffusion equations has been conducted by \citet{Dawson1993} in \citeyear{Dawson1993}. The group investigated the classical Selkov model (originating from biological studies of glycolysis).
They used a hexagonal lattice with a BGK relaxation model and an explicit first-order integration of the source term. 
The proposed technique was used to simulate pure diffusion, homogenous reaction and the formation of Turing patterns. 
Later, the model was re-formulated on a square lattice by \citet{blaak2000lattice}. 
The first order approximation of the source term was present in both works.

Arguably, the most widely explored application of the advection-diffusion \ac{LBM} is found in coupled flow-heat transfer problems  \cite{Doolen1998,Guo2007,Karlin2013,Fei2018a,Fei2018c,Premnath2018cascadedSources}.
To model the underlying physics, a two-population approach is frequently used.
One of the populations simulates the motion of the fluid, while the second one uses some form of the energy balance equation to resolve the heat transfer.
The primary benefit of using the same numerical framework for fluid flow and heat transfer is that it greatly simplifies the computational implementation.

The \ac{LBM} framework has also been extended to capture the conjugate heat transfer and or phase change between solid and liquid phases.
Huang et al. \cite{Huang2013,Huang2015a} incorporated the term responsible for the latent heat of fusion into the equilibrium distribution function. 
As a result, moments of equilibrium have been adjusted to recover the desired \ac{PDE}.
A similar effect has been obtained by \citet{Hosseini2018}, who presented a set of weight coefficients to tune the second-order moments of a source-like term, which was added to the equilibrium distribution function.
\citet{Karani2015}, and later \citet{Chen2017b}, studied the macroscopic energy equation suitable to model conjugate heat transfer in comparison with the standard \ac{ADE}. 
The difference was expressed as a corrective source term, and was added to the standard \ac{LBM} routine for the \ac{ADRE} to obtain the correct energy balance.
While the aforementioned works \cite{Karani2015,Chen2017b,Hosseini2018} introduce a source term specific for conjugate heat transfer problems, 
they do not discuss its treatment which potentially leads to the use of explicit first-order schemes.
Interested readers are referred to a comprehensive review of methods related to phase-change, heat transfer and multiphase flows by \citet{Li2016}.

The source term technique can be also used to implement a boundary condition on a moving or curved wall.  
The intensity of the heat source is adjusted in each iteration to match the Dirichlet, Neumann, or Robin boundary condition.
Models that use this technique are commonly referred to as \acl{IBM}.
\citet{Seta2013} performed a comprehensive work on its thermal variant, including an analysis of the error terms present.
Many authors have recognised the importance of consistently treating the source term with an appropriate scheme \cite{Jeong2010,Kang2011,Kang2011_thermalIBM,Seta2013,Huang2014,Hu2015,Wang2020}, however, it is not always taken into account~\cite{Eshghinejadfard2016,Karimnejad2019,Suzuki2018}.
An alternative approach to couple the fluid-solid interaction is through the partially saturated method \cite{noble1998lattice}, where a lattice node is categorised as a pure fluid, pure solid or mixed one.

To simulate advection-diffusion-reaction in a compressible medium, \citet{Aursjo2017} have defined the concentration of the scalar quantity relative to the density of fluid.
Later, the same research group proposed a scheme for including a mass source term in the \ac{NS} equations and showed its application for imposing a pressure boundary condition~\cite{Aursjo2018}.

While problems relating to heat transfer represent the majority of work conducted with the \ac{ADE} resolved in the \ac{LBM} framework, other physical phenomena are also studied. 
When multiple components are present in the same volume, a reaction between them may occur \cite{Dawson1993,kang2006lattice,Lin2016,Hosseini2020}.
\citet{kang2006lattice} managed to reduce the number of equations required to solve the problem of dividing species by separating their interaction into reaction rate-, and diffusion-dominated.
Again, the convergence was not discussed in \cite{Dawson1993,kang2006lattice,Lin2016}.

\citet{Shi2009} and \citet{Chai2016} provided general studies on the non-linear and anisotropic variants of the advection-diffusion equations in the \ac{LBM} framework, respectively. 
The influence of source term treatment is highlighted, but the numerical examples are focused on other aspects of the \ac{ADRE} and do not include an example of solving non-linear source term preserving both temporal and spatial derivatives. 
The asymptotic analysis of the \ac{LBM} schemes for the \ac{ADE} have been discussed in detail by~\citet{Yoshida2010} and~\citet{Chai2020}. 
The authors gave a detailed expansion analysis for both \ac{SRT} and \ac{MRT} collision operators for the advection-diffusion-reaction system.  
For the diffusive scaling, second-order accuracy was reported. However, the test cases presented did not include source terms, which were integrated with a first-order scheme.

Having outlined the range of applications, the mathematical methodology is now presented. 
Readers interested in a detailed discussion on analysis methods are referred to~\cite{Chai2020,Fucik2021,SSimonis_2023}.
In general, the derivations of the conventional \ac{LBM} numerical schemes present in the literature, can be broadly divided into two groups.
First one is referred as \textit{bottom-up} and the second one as \textit{top-down} approach.

The \textit{bottom-up} procedure starts from an \textit{a priori} postulated discrete evolution scheme. 
It proceeds with an expansion procedure (such as Chapman-Enskog) to recover the macroscopic equations and to analyse its order of accuracy.
Knowing the difference between the target and recovered equation, one can apply corrections to account for the missing terms.
Regarding the advection-diffusion-reaction equation, the missing terms are recognised as spatio-temporal derivatives of the source term.
To regain second-order accuracy of the numerical scheme, two approaches can be distinguished.
Shi et al.~\cite{Shi2008,Shi2009} evaluated the derivatives using standard finite-difference stencils.
On the other hand,~\citet{Seta2013} and~\citet{Chai2016} realised that a redefinition of variables allowed the problem to be resolved, while preserving the locality of the underlying scheme.
An analogous analysis regarding the forcing term in the \ac{NS} equations have been done by~\citet{Guo2002}.

In the \textit{top-down} approach, the \ac{DBE} is first constructed.
It is then integrated along its characteristics to derive the fully discrete scheme~\cite{he1998discrete,lee2005astable,guo2018general}. 
In most cases, the resulting time integration scheme will be implicit.
Again, redefinition of variables can be used to transform it into explicit equations~\cite{Doolen1998}.
Finally, it is interesting to notice, that some alternative \ac{LBM} schemes are also being developed~\cite{Kramer2018,Wilde2019,Strzelczyk2022}.

\subsection{Structure}
With the state of the art, and the common methodologies for deriving an \ac{LBM} scheme introduced, the remainder of this paper is structured as follows.
In \cref{sec:DBE}, a general framework based on a \textit{top-down} approach is discussed in detail.
From this, a general method which preserves the second-order nature of the underlying \ac{LBM} with source terms is analysed. 
A simplified relaxation procedure using a moment-based collision operator is then presented in \cref{sec:the_collision_operator}.
\Cref{sec:acoustic_diffusive_scalling} outlines the non-dimensionalisation procedure for a PDE and the required scaling of parameters on \ac{LBM} grids.
Subsequently,~\cref{sec:linear_ADRE} describes how leading components of error affect the accuracy of a solution in terms of both spatial and temporal resolution.
To illustrate the methodology, the proposed approach for solving a formally implicit source term is applied to the Allen-Cahn equation in \cref{sec:AllenCahn}. 
With the use of symbolic algebra, a solution with an explicit algorithm, which preserves second-order accuracy is proposed.
Finally, the methodology is validated numerically by comparison with a finite-element method solution in \cref{sec:fem_lbm} and a convergence study is provided in \cref{sec:AllenCahn_uniform_IC,sec:periodic_2D_Da_convergence}. 
The summary of the findings and major outcomes of this work are presented in \cref{sec:conclusions}.

\noindent\fbox
{
    \parbox{\textwidth}
    {\printnomenclature}
}

\section{Model description}
\subsection{Discrete Boltzmann equation (\acs{DBE}) }\label{sec:DBE}

Here, the iterative scheme is derived using the \textit{top-down} approach, i.e. by direct integration of the \ac{DBE}.
For a set of densities $\h_i(\bs x, t)$, and velocity vectors $\mathbf{e_i}$, the \ac{DBE} is known as (\cite[Chapter~3.4 and~8.3]{Kruger2017}),
\begin{linenomath}\begin{align}
\pr{t}\h_i + e^j_i \pr{x_j}\h_i &= \dfrac{1}{\tau} \left(\h^\eq_i(\phi, \bs u)-\h_i \right) + q_i(\phi, \bs x,t), \label{eq:DBE}
\end{align}\end{linenomath}
where $\h^\eq$ is the equilibrium distribution, $q$ is the source term, $\phi=\sum_i\h_i$ is the scalar field, and $\bs u$ is the advection velocity.
It is important to reiterate here that the source term depends on the investigated scalar field. 
With an appropriate choice of equilibrium distribution~\cite{Kruger2017}, this equation can be shown to represent the advection-diffusion-reaction equation with, $\phi=\sum_i\h_i$, and $Q=\sum q_i$.
The equations for the equilibrium distribution are presented in~\cref{sec:moments_of_DF}.
An important property of the equilibrium that will be used, is, 
\begin{linenomath}\begin{equation}
\sum_i\h^\eq_i(\phi, \bs u) = \phi.
\end{equation}\end{linenomath}

Contrary to classical numerical methods, such as finite-volume or finite-element method, 
the space and time integration \underline{can not} be treated independently in the construction of a conventional \ac{LBM} scheme.
For a fixed $i$, $x$ and $t$, the characteristic of \ac{DBE} is given by $\bs x(s) = \bs x + s\,\bs e_i$, and $t(s) = t + s$.
Integrating over $s$ from 0 to 1 we get,
\begin{linenomath}\begin{align}
\underbrace{ \int_0^1 \left( \pr{t}\h_i +  e^j_i \pr{x_j} \h_i\right) ds}_{I_1} 
&= 
\underbrace{\int_0^1 \left(\dfrac{1}{\tau}\left(\h^\eq_i-\h_i\right) + q_i\right)ds}_{I_2}. \label{eq:DBE_with_source_term}
\end{align}\end{linenomath}
Two integrals, $I_{1}$ and $I_{2}$, can be identified in the integration process of the \ac{DBE}.
To denote the variables from the next time step, the hat superscript is used in this work. 
Namely, $\hat{\bs x} =\bs x + \bs e_i$, $\hat{t} = t+1$, $\hat\phi=\phi(\hat x,\hat t)$ and $\hat{\bs u} = \bs u (\hat x, \hat t)$.
The first integral can be evaluated directly, as it is a material derivative of $h_i$,
\begin{linenomath}\begin{align}
I_1 &=\int^{1}_{0} \left( \pr{t}\h_i +  e^j_i \pr{x_j} \h_i \right) ds =  h_i(\hat{\bs x},\hat{t}\,) - h_i(\bs x,t).
\end{align}\end{linenomath}
The second integral can be approximated by the trapezoidal rule,
\begin{linenomath}\begin{align}
I_2 &= \int_0^1 \left(\dfrac{1}{\tau}\left(\h^\eq_i-\h_i\right) + q_i\right)ds \nonumber \\
 & \simeq  \dfrac{1}{2}  \left[
 \dfrac{1}{\tau}\left( \h^\eq_i(\hat{\phi},\hat{\bs u})-\h_i(\hat{\bs x},\hat{t})\right) + q_i(\hat{\phi}, \hat{\bs x}, \hat{t})
 + \dfrac{1}{\tau}\left( \h^\eq_i(\phi, \bs u)-\h_i(\bs x,t)\right) + q_i(\phi, \bs x, t)
\right]. \label{eq:trapezoidal_rule}
\end{align}\end{linenomath}

Integrating \cref{eq:trapezoidal_rule} over the characteristics, 
and collecting the variables from the next time step (which depend on $\hat{\bs x}$ and $\hat{t}$) on the left hand side gives, 
\begin{linenomath}\begin{align}
\underbrace{\left[ 1 + \dfrac{1}{2 \tau} \right] \h_i(\hat{\bs x},\hat{t}\,) - \dfrac{1}{2\tau}\h^\eq_i(\hat{\phi},\hat{\bs u}) - \dfrac{1}{2} q_i(\hat{\phi}, \hat{\bs x}, \hat{t} \,)}_{\text{~\cref{eq:1defining_tilde_f}}} = 
\left[ 1- \dfrac{1}{2 \tau} \right] \h_i(\bs x,t\,)  + \dfrac{1}{2\tau}\h^\eq_i(\phi, \bs u) + \dfrac{1}{2} q_i(\phi, \bs x, t).  \label{eq:before_defining_tilde_f}
\end{align}\end{linenomath}
Next, a shifted distribution, $\lbmh$, is introduced to remove the implicit relation from \cref{eq:before_defining_tilde_f}.
To illustrate the idea, a simple example of variable shift in a ordinary differential equation is presented in~\ref{app:ODEexample}.
A general description of this procedure in the contexts of \ac{LBM} can be found in the textbook by \citet[Chapter~3.5]{Kruger2017}).

The new (shifted) distribution, denoted with a tilde, is defined as,
\begin{linenomath}\begin{align}
 \lbmh_i (\bullet) &= \left[ 1 + \dfrac{1}{2 \tau} \right] h_i(\bullet) - \dfrac{1}{2\tau}h^{eq}_i(\bullet) - \dfrac{1}{2} q_i(\bullet) \label{eq:1defining_tilde_f} \\
\implies h_i(\bullet)&= \frac{1}{1 + \frac{1}{2 \tau}}\left( \lbmh_i(\bullet) + \dfrac{1}{2\tau}h^{eq}_i(\bullet) + \dfrac{1}{2} q_i(\bullet)\right), \label{eq:2defining_tilde_f}
\end{align}\end{linenomath}
where $\bullet$ is a placeholder for variables in either $t$ or $\hat{t}$.
Substituting \cref{eq:1defining_tilde_f} into the left-hand side of~\cref{eq:before_defining_tilde_f} 
and~\cref{eq:2defining_tilde_f} into the right-hand side of~\cref{eq:before_defining_tilde_f} 
leads to a fully explicit evolution scheme (see \ref{app:derivations_DBE} for details),
\begin{linenomath}\begin{align}
\lbmh_i(\bs x + \bs e_i,t + 1) = 
\lbmh^\star_i(\bs x,t) = 
(1 - \omega) \lbmh_i(\bs x,t) + \omega \h^\eq_i(\phi, \bs u) + \left(1 - \dfrac{\omega}{2}\right)q_i(\phi, \bs x, t), \label{eq:tilde_f_evolution}
\end{align}\end{linenomath}
where a relaxation frequency, $\omega=\dfrac{1}{\tau +1/2}$, has been introduced to simplify the expression.
The post-collision densities are denoted as $\lbmh^\star$.
It is important to underline, that $\lbmh$ is the variable solved in the implementation of a \ac{LBM} scheme, and is denoted with a tilde in this paper to distinguish it from the non-shifted population, $\h$.
The differences in the discretisation schemes available in the literature are discussed in 
\ref{app:comparison_of_approaches}.

\subsection{Calculation of the scalar field}\label{sec:phi_from_phi_table}
\Cref{eq:tilde_f_evolution} would be an explicit iterative scheme for $\lbmh$, if not for the implicit dependence on the scalar field, $\phi$.
One can calculate $\phi$ from $\lbmh$ by summation of ~\cref{eq:1defining_tilde_f} noting that both $\sum_i\h_i$ and $\sum_i\h^\eq_i$ are equal to $\phi$,
\begin{linenomath}\begin{align}
\tilde \phi &= \sum_i\lbmh_i = \left( 1 + \dfrac{1}{2 \tau} \right) \sum_i\h_i (\bs x,t)- \dfrac{1}{2\tau}\sum_i\h^\eq_i(\phi, \bs u) - \dfrac{1}{2} \sum_i q_i(\phi, \bs x,t) \nonumber \\
 &= \left( 1 + \dfrac{1}{2 \tau} \right) \phi -  \dfrac{1}{2 \tau} \phi -  \dfrac{1}{2} Q(\phi, \bs x,t) \nonumber \\
 &= \phi - \dfrac{1}{2}Q(\phi, \bs x,t). \label{eq:tilde_phi_from_phi}
\end{align}\end{linenomath}
Since the source term depends on the scalar field, $Q=Q(\phi, \bs x,t)$, one must solve the implicit equation for $\phi$ to retain the second-order convergence of the underlying scheme. 
This can be done with a sub-iteration routine, such as bisection or Newton's method, and in certain cases it can be solved analytically.
In \cref{tab:phi_solutions}, a set of common source terms is presented with corresponding equations for the scalar field, $\phi$.
In some cases multiple solutions to the implicit equation can exist, and physical limitations must be applied to select the appropriate value. 
For these cases, the relevant conditions are listed in the notes.
If $Q$ is dependent only on the spatial , $\bs x$, and temporal, $t$, location, the equation reduces to, $\phi = \tilde\phi + \frac{1}{2}Q(\bs x,t)$, which is commonly encountered in the literature.

\begin{table}[htbp]\centering
\caption{Calculation of the value of the macroscopic scalar field $\phi$ from the sum of \ac{LBM} densities $\tilde\phi=\sum_i\lbmh_i$. In cases of higher order terms,  only one branch of the implicit function is selected, based o physical considerations.}\label{tab:phi_solutions}
\begin{tabular}{lllp{3.7cm}}
\toprule
Class & Source term & Equation for $\phi$ & Notes\\
\midrule
No source term & $Q = 0$ & $\phi = \tilde\phi$\\
$\phi$ independent term & $Q = Q(x,t)$ & $\phi = \tilde\phi + \frac{1}{2}Q(x,t)$\\[1ex]
General 1st order term & $Q = -\lambda(\phi - \gamma)$ & $\phi = \tilde\phi + \frac{1}{2 + \lambda}\lambda(\gamma - \tilde\phi)$\\[1ex]
Linear decay & $Q = -\lambda\phi$ & $\phi = \frac{2}{2 + \lambda}\tilde\phi$\\[1ex]
General 2nd order term & $Q = -\lambda(\phi^2 - B\phi+C)$ & $\phi = \alpha + \sqrt{\frac{2}{\lambda}\tilde\phi + \alpha^2 - C}$ & for $\phi>\alpha$, where $\alpha = \frac{B}{2} - \frac{1}{\lambda}$\\[1ex]
Logistic model & $Q = \lambda\phi(1-\frac{\phi}{\gamma})$ & $\phi = \alpha + \sqrt{\frac{2\gamma}{\lambda}\tilde\phi + \alpha^2}$ & for $\phi>\alpha$, where $\alpha = \gamma\left(\frac{1}{2} - \frac{1}{\lambda}\right)$\\[1ex]
Gompertz model & $Q = -\lambda\phi\ln{\frac{\phi}{\gamma}}$ & $\phi = \alpha e^{\mathcal{W}_0(\frac{2}{\lambda}\frac{\tilde\phi}{\alpha})}$ & for $\phi > \alpha\frac{1}{e}$, where $\alpha = \gamma e^{-\frac{2}{\lambda}}$, and $\mathcal{W}$ is the Lambert W function\\[1ex]
Allen-Cahn equation & $Q = \lambda \phi \left(1-\phi^2\right)$ & $\phi = \frac{A}{C} - C$ & for $\lambda<2$, where $A=\frac{2-\lambda}{3\lambda}$, $B = \frac{\tilde\phi}{\lambda}$ , and $C = \sqrt[3]{\sqrt{B^2 + A^3} - B}$\\[1ex]
General term & $Q = Q(\phi,x,t)$ & $\phi - \frac{1}{2}Q(\phi,x,t) = \tilde\phi$ & implicit equation, which can be solved with sub-iterations \\
\bottomrule
\end{tabular}
\end{table}

Observe, that the equation for the macroscopic quantity, $\phi$, is independent of the \ac{LBM} scheme being used.
Namely, for any collision operator, for example \ac{SRT}, \ac{MRT} or cascaded LBM, and any discretisation of the source term, $q_i$, the equation will stay the same.

\subsection{Initialization}
As the densities simulated in the \ac{LBM} are different than the \ac{DBE} densities, 
the difference must to be taken into account in the initialization procedure.
Namely, if one assumes the \ac{DBE} to be initialized by the equilibrium distribution function, the \ac{LBM} densities need to be initialized as,
\begin{linenomath}\begin{align}
\lbmh_i (x,0) &= h^{eq}_i(\phi_0,\bs u_0) - \dfrac{1}{2} q_i(\phi_0, x, 0) \nonumber \\
 &= h^{eq}_i(\tilde{\phi}_0,\bs u_0) 
,\label{eq:initialization}
\end{align}\end{linenomath}
where $\phi_0$ and $\bs u_0$ are the initial values of the macroscopic field and advection velocity respectively.
Taking advantage of~\cref{eq:tilde_phi_from_phi}, the initialization procedure can be simplified by calculating $\tilde{\phi}_0$.
Omission of this crucial adjustment of the initialization procedure will result in a solution inconsistent with the desired initial condition, and will reduce the order of convergence. 
Note, that~\cref{eq:initialization} gives the value of the initial distribution functions after streaming and before collision.

\subsection{Moments of the distribution}\label{sec:moments_of_DF}
In order to formally discuss different \ac{LBM} schemes, the concept of moments of the distribution is introduced in this section.
Here, the D2Q9 lattice will be adopted for illustrational purposes, but without the loss of generality. 
The D2Q9 is the most popular lattice discretisation used for 2D problems, and it defines the discrete velocity vectors $\bs e_i$ as,
\begin{eqnarray*}
{\bs e}_1 = [0,0] & {\bs e}_2 = [1,0] & {\bs e}_3 = [0,1] \\
{\bs e}_4 = [-1,0] & {\bs e}_5 = [0,-1] & {\bs e}_6 = [1,1] \\
{\bs e}_7 = [-1,1] & {\bs e}_8 = [-1,-1] & {\bs e}_9 = [1,-1]
\end{eqnarray*}
The discrete, raw moments in 2D are defined as,
\begin{linenomath}\begin{align} 
\Upsilon_{mn} &= \sum_i(e^x_i)^m (e^y_i)^n \h_i, \label{eq:discrete_raw_moments_definition}
\end{align}\end{linenomath}
which can be easily extended to 3D with three indices.
For moments of \ac{LBM} densities, $\lbmh$, post-collision densities, $\lbmh_i^\star$, equilibrium distribution, $\h^\eq$ and source term, $q$, we will use $\tilde\bY$, $\tilde\bY^\star$, $\bY^\eq$ and $\bs R$ respectively.

As the set of densities is finite, and the velocities $\bs e$ are distinct, 
one can select a finite set of linearly independent moments to fully represent any set of densities.
For the D2Q9 lattice, the following set of moments is chosen,
\begin{linenomath}\begin{equation}
\bY = 
\begin{bmatrix}
\Upsilon_{00}& \Upsilon_{10}& \Upsilon_{01}& \Upsilon_{20}& \Upsilon_{02}& \Upsilon_{11}& \Upsilon_{21}& \Upsilon_{12}& \Upsilon_{22}\end{bmatrix}^\top.
\end{equation}\end{linenomath}

The transformation from the densities, $\h$, to the vector of moments can be expressed through matrix multiplication as $\bY = \mathbb{M} \boldsymbol{\h} $. 
In the case of D2Q9, this matrix reads,
\begin{linenomath}\begin{align} 
\mathbb{M}&=
\begin{bmatrix}
     1  &  1 &  1 &  1 &  1 &  1  &  1 &  1 &  1\\
     0  &  1 &  0 & -1 &  0 &  1  & -1 & -1 &  1\\
     0  &  0 &  1 &  0 & -1 &  1  &  1 & -1 & -1\\
     0  &  1 &  0 &  1 &  0 &  1  &  1 &  1 &  1\\
     0  &  0 &  1 &  0 &  1 &  1  &  1 &  1 &  1\\
     0  &  0 &  0 &  0 &  0 &  1  & -1 &  1 & -1\\
     0  &  0 &  0 &  0 &  0 &  1  &  1 & -1 & -1\\
     0  &  0 &  0 &  0 &  0 &  1  & -1 & -1 &  1\\
     0  &  0 &  0 &  0 &  0 &  1  &  1 &  1 &  1\\
\end{bmatrix}.\label{eq:M_matrix}
\end{align}\end{linenomath}

It is convenient to express \ac{LBM} schemes in terms of moments, and not densities themselves.
This is because they represent specific mechanics, and or have physical interpretations, which the densities themselves do not possess.
The \ac{LBM} collision operator defined in equation~\cref{eq:tilde_f_evolution}, can be expressed using moments as,
\begin{linenomath}\begin{equation}
\tilde\bY^\star(\bs x, t) = (1 - \omega) \tilde\bY(\bs x, t) + \omega \bY^\eq(\phi, \bs u) + \bigg(1 - \dfrac{\omega}{2}\bigg)\bs R(\phi, \bs u, \bs x, t).
\end{equation}\end{linenomath}
The moments of the discrete equilibrium, $\bY^\eq$, are defined to be equal to moments of the continuous Maxwell-Boltzmann distribution function, $\bY^\eq(\phi,\bs u) = \phi\,\bs\Gamma(\bs u)$, where,
\begin{linenomath}\begin{align}
\bs\Gamma(\bs u)=\begin{bmatrix}
     1 &
     u_x &
     u_y &
     c_s^2 + u_x^2 &
     c_s^2 + u_y^2 &
     u_x u_y &
     u_y(c_s^2 + u_x^2) &
     u_x(c_s^2 + u_y^2) &
     c_s^4 + c_s^2(u_x^2+u_y^2) + u_x^2 u_y^2
\end{bmatrix}^\top.  \label{eq:Gamma}
\end{align}\end{linenomath}
One can clearly see that these moments fulfil the conditions required for representation of the advection-diffusion-reaction equation by the \ac{DBE}.
A common choice of the speed of sound $c_s$ is $1/\sqrt{3}$, which reduces the error in higher moments.
The moments of the equilibrium distribution are truncated by some authors on the second order terms 
\cite{Doolen1998,Guo2007,Huang2013,Karlin2013} 
, by not including the terms $u_x^2 u_y$, $u_x u_y^2$ and $u_x^2 u_y^2$.
Furthermore, it can be truncated at first-order terms 
\cite{Yoshida2010,Karani2015,Chen2017b,Hosseini2018} 
giving,
\begin{linenomath}\begin{equation}
\bs\Gamma^\text{1st order}(\bs u)=\begin{bmatrix}
     1 &
     u_x &
     u_y &
     c_s^2 &
     c_s^2 &
     0 &
     u_y c_s^2 &
     u_x c_s^2 &
     c_s^4
\end{bmatrix}^\top.
\end{equation}\end{linenomath}
Formulas analogous to \cref{eq:Gamma} can be used for the D2Q5 lattice~\cite{Fei2018c}.
Similarly, the moments of the source term can be expressed as $\bs R=Q(\phi, \bs x,t) \,\bs\Gamma(\bs u)$. 
Again, some authors simplify the source term by truncating the velocity \cite{Wang2007,Yoshida2010}.
Investigation of an error related to the order of velocity expansion of the equilibrium distribution function has been conducted by \citet{Chopard2009}. 
It has been shown that the full order velocity expansion enhances both the Galilean invariance \cite{Geier2006,Nie2008,Fei2018a,DeRosisLuo2019} 
and stability of the \ac{LBM} schemes \cite{Geier2006,DeRosisLuo2019}.

The equilibrium and source term distribution functions can be easily calculated from their moments as $\boldsymbol{h}^{eq} = \mathbb{M}^{-1} \bY^{eq}$ and $\boldsymbol{q} = \mathbb{M}^{-1} \boldsymbol{R}$.

\subsubsection{The collision operator}\label{sec:the_collision_operator}
If the moments of the source term are chosen to be $\bs R=Q(\phi, \bs x,t) \,\bs\Gamma(\bs u)$, and $\bY^\eq(\phi,\bs u) = \phi\,\Gamma(\bs u)$, 
one can use \cref{eq:tilde_phi_from_phi,eq:Gamma} to express the collision operator in three equivalent forms,
\begin{linenomath}\begin{align}
\tilde\bY^\star &= (1 - \omega) \tilde\bY + \omega \bY^\eq(\phi,\bs u) + \bigg(1 - \dfrac{\omega}{2}\bigg)\bs R\\
&= (1 - \omega) \tilde\bY + \omega \bY^\eq(\tilde\phi,\bs u) + \bs R\label{eq:almost_ultimate_tilde_f_evolution}\\
&= (1 - \omega) \bigg(\tilde\bY - \bY^\eq(\tilde\phi,\bs u)\bigg) + \bY^\eq(\tilde\phi + Q,\bs u).\label{eq:EDM_f_evolution}
\end{align}\end{linenomath}
The form provided in \cref{eq:almost_ultimate_tilde_f_evolution} can be viewed as a simplification of a relaxation procedure expressed in central moments space by \citet{Fei2018c}. 
On the other hand, the form in \cref{eq:EDM_f_evolution} can be interpreted as a natural extension of the exact difference method~\cite{Kupershtokh2009} for reaction-type source terms. 
Even though~\cref{eq:almost_ultimate_tilde_f_evolution,eq:EDM_f_evolution} use $\bY^\eq(\tilde\phi,\bs u)$ instead of $\bY^\eq(\phi,\bs u)$, the implicit~\cref{eq:tilde_phi_from_phi} must still be resolved as the source term $Q$ is dependent on $\phi$.

For a comparison of forcing schemes, the reader is referred to the studies~\cite{Guo2002,fei2017consistent}.  
Further discussion concerning discretisation and the order of the velocity expansion of the forcing scheme in the central moment space can be found in~\cite{Huang2018,DeRosis2019,gruszczynski2020cascaded} and references therein.

\subsubsection{Two relaxation time}

The \acf{TRT} collision operator was shown to have superior accuracy and 
stability~\cite{Ginzburg2005,Kuzmin2011b}
when compared to the \ac{SRT} collision presented in the previous section.
The \ac{TRT} approach consists of decomposing the distribution function into symmetric and anti-symmetric components~\cite{Ginzburg2005}, 
\begin{subequations}
\begin{linenomath}\begin{align}
\h_j^{even}&=\frac{\h_j + \h_k}{2}, \\
\h_j^{odd}&=\frac{\h_j - \h_k}{2},
\end{align}\end{linenomath}
\end{subequations}
for $k$ chosen such that $\bs e_j = -\bs e_k$. For the two components, two different relaxation times, $\tau$, are used resulting in two different relaxation coefficients, $\omega$.

The symmetric component, $\h_j^{even}$, has only even order moments, and the anti-symmetric component has only odd order moments that are non-zero.
This means that a \ac{TRT} collision can be constructed from~\cref{eq:almost_ultimate_tilde_f_evolution} as,
\begin{linenomath}\begin{align}
\tilde\bY^\star
&= 
(1 - \mathbb{S}) \tilde\bY
+ \mathbb{S} \bY^\eq(\tilde\phi,\bs u)
+ \boldsymbol{R}, \label{eq:ultimate_tilde_f_evolution_trt}
\end{align}\end{linenomath}
where $\mathbb{S}$ is the diagonal relaxation matrix, which has $\omega_{odd}$ in rows corresponding to moments of odd order, and $\omega_{even}$ otherwise.
For D2Q9 the matrix is as follows,
\begin{linenomath}\begin{align}
\mathbb{S} = diag \left( [\omega_{even}, \omega_{odd}, \omega_{odd}, \omega_{even}, \omega_{even},\omega_{even}, \omega_{odd}, \omega_{odd}, \omega_{even}] \right).
\end{align}\end{linenomath}
This means that the non-orthogonal matrix of raw moments, $\mathbb{M}$, provided in \cref{eq:M_matrix}, diagonalises the \ac{TRT} collision operator \cite{Asinari2008}.
For additional information on the \ac{TRT} method, the interested reader is directed to the work of \citet{Ginzburg2005}, although this particular property has not been explicitly stated there.
In case of the \ac{ADE}, the relaxation rate for odd moments, $\omega_{odd}$, has to correspond to the macroscopic diffusion coefficient,
\begin{linenomath}\begin{align}
\omega_{odd} = \frac{1}{\frac{M}{c_s^2} + 1/2}.
\end{align}\end{linenomath}
On the other hand, the (tunable) relaxation rate for even moments, $\omega_{even}$, can be defined based on the so-called magic parameter,
\begin{linenomath}\begin{align}
\Lambda = \left( \dfrac{1}{\omega_{odd}} - \frac{1}{2} \right) \left( \dfrac{1}{\omega_{even}} - \frac{1}{2} \right).
\end{align}\end{linenomath}
Fixing the magic parameter at different constant values, results in a minimisation of specific types of discretisation errors, and can improve stability and or accuracy of commonly used boundary conditions~\cite{Ginzburg2005,Kuzmin2011b}.
\subsubsection{Streaming}
As usual, the collision step, given by \cref{eq:ultimate_tilde_f_evolution_trt}, is followed by the streaming step,
\begin{linenomath}\begin{align}
\lbmh_i(\bs x + \bs e_i,t + 1) = 
\lbmh^\star_i(\bs x,t) = 
\mathbb{M}^{-1} \tilde\bY^\star(\bs x,t). \label{eq:Mstreaming}
\end{align}\end{linenomath}
Concluding, the order of operations in the \ac{LBM} scheme can be presented as in~\cref{fig:CalculationFlow}.

\tikzstyle{block} = [rectangle, draw, fill=gray!15, node distance=15em,
    text width=11em, text centered, rounded corners, minimum height=5em]
\tikzstyle{line} = [draw, -latex']
\begin{center}
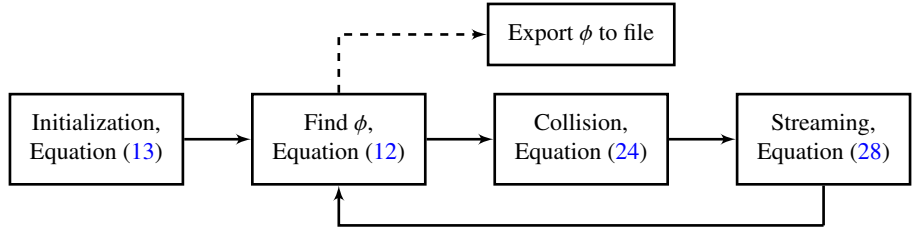
 
\begin{tikzpicture}[
	box/.style={ shape=rectangle, draw, align=center, inner sep = 0.75em, font=\small},
	point/.style={coordinate},>=stealth', line width=1pt]
    \node [box] (init)  {Initialization, \\ \cref{eq:initialization}};
    \node [box, anchor=west] (CalcPhi) at ([xshift=2.5em] init.east) {Find $\phi$, \\ \cref{eq:tilde_phi_from_phi}};    
    \node [box, anchor=west] (collide) at ([xshift=2.5em] CalcPhi.east) {Collision, \\ \cref{eq:ultimate_tilde_f_evolution_trt} };
    \node [box, anchor=west] (stream) at ([xshift=2.5em] collide.east) {Streaming, \\ \cref{eq:Mstreaming}};
    \node [box, anchor=south] (save) at ([yshift=1em] collide.north) {Export  $\phi$ to file};
    \node [point] (p1) at ([yshift=-1.5em] stream.south) {};
    \path [line]  (init) -- (CalcPhi);
    \path [line]  (CalcPhi) -- (collide);
    \path [line]  (collide) -- (stream);
    \path [line, dashed]  (CalcPhi) |- (save);
    \path [line]  (p1) -| (CalcPhi);
    \draw [] (stream) -- (p1); 
\end{tikzpicture}
\captionsetup[figure]{name=Figure}
\captionof{figure}{The order of operations performed during execution of the solver.}
\label{fig:CalculationFlow}
\end{center}

\section{Scaling of LBM}
\label{sec:acoustic_diffusive_scalling}
To represent the same physical problem, the non-dimensional form of the investigated differential equation must be preserved on a set of \ac{LBM} grids.
Let us consider a source term $Q = \lambda P(\phi)$, where $P$ has the same unit as $\phi$, and $\lambda$ is a scaling coefficient (see Table~\ref{tab:phi_solutions}).
Using $L$, $T$ and $U$ as reference length, time and velocity, respectively, one can define non-dimensional coordinates $x=x^\ast L$, $t=t^\ast T$, $u=u^\ast U$.
For a constant $M$ one can express the \cref{eq:ADE} in non-dimensional form~\cite{Langtangen2016},
\begin{linenomath}\begin{align}
\pr{t^\ast}\phi+ \text{\acs{Pe}}\,\text{\acs{Fo}}\, \nabla^\ast\cdot(\boldsymbol{u^\ast}\phi)=
\text{\acs{Fo}}\, \Delta^\ast \phi+ \text{\acs{Da}}\,\text{\acs{Fo}}\, P(\phi),
\end{align}\end{linenomath}
where \acs{Fo}, \acs{Da} and \acs{Pe} are dimensionless numbers:
\begin{itemize}
\item \acl{Fo} $\text{\acs{Fo}}= \frac{MT}{L^2}$ --- the ratio of the diffusive term to the temporal term;
\item (second) \acl{Da} $\text{\acs{Da}}= \frac{\lambda L^{2}}{M}$ --- the ratio of the reaction term to the diffusive term;
\item \acl{Pe} $\text{\acs{Pe}}= \frac{U L}{M}$ --- the ratio of the convective term to the diffusive term.
\end{itemize}
Observe, that the \ac{Fo} incorporates the temporal scale of the simulation and as such can be considered as a non-dimensional time.
An alternative approach would be to fix $\text{\acs{Fo}}=1$, and quote the time of simulation for each case.

The \ac{LBM} grids are described by their characteristic length and time $\left(L,T\right)$ expressed in number of elements and time-steps.
The element size, and time-step can be expressed as their inverse, $\Delta x=\frac{1}{L}$ and $\Delta t=\frac{1}{T}$.
For each grid, there are corresponding values of  $M$, $\lambda$, and $U$ that preserve the dimensionless numbers: $M = \text{\acs{Fo}}\frac{L^2}{T}$, $\lambda = \text{\acs{Da}}\, \text{\acs{Fo}}\frac{1}{T}$ and $U = \text{\acs{Pe}}\, \text{\acs{Fo}}\frac{L}{T}$.

For the so-called acoustic scaling, one uses a series of lattices with $L_k=\varepsilon_k^{-1} L_0$ and $T_k=\varepsilon_k^{-1} T_0$, for some scaling factor $\varepsilon_k\to 0$.
Using the previously mentioned equations for the simulation parameters one obtains, $M_k=\varepsilon_k^{-1} M_0$, $U_k=U_0$, and $\lambda_k = \varepsilon_k\lambda_0$.
This scaling is called acoustic, as it preserves the velocity scale, making the speed of sound constant across the series of \ac{LBM} grids.
On the other hand, for diffusive scaling, one has $L_k=\varepsilon_k^{-1} L_0$ and $T_k=\varepsilon_k^{-2} T_0$.
This in turn gives, $M_k= M_0$, $U_k=\varepsilon_k U_0$, and $\lambda_k = \varepsilon_k^{2}\lambda_0$.
This scaling is called diffusive, as the diffusion coefficient $M$ is constant across the series of \ac{LBM} grids.

\section{Model Verification and Validation} \label{sec:model_verification_and_validation}
To test the numerical properties of the described \ac{LBM} scheme, two equations were investigated, 
an advection diffusion equation, with a linear source term, and the Allen-Cahn equation.
As analytical solutions are easy to obtain for the first example, a detailed analysis of error is performed in that case.
Also, for the first equation, the results are compared with the situation in which one would naively assume $\phi=\tilde\phi$.
In the case of the Allen-Cahn equation, which has a bi-stable, highly non-linear source term, convergence is checked for both uniform, and non-trivial initial conditions.

If not stated differently, the error between the numerical solution $\phi(x_i)$ defined on the lattice points $\bs x_i$ 
and the reference solution $\phi_\text{ref}$ is quantified using the $\mathcal{L}_2$ norm, defined as:
\begin{linenomath}\begin{align}
\mathcal{L}_2 \text{ norm of error} = \sqrt{\frac{1}{N}\sum_{i=1}^{N}\left(\phi(x_i)-\phi_\text{ref}(x_i)\right)^2},
\end{align}\end{linenomath}
where $N$ is the number of points in the lattice.

\subsection{Linear advection-diffusion-reaction model}\label{sec:linear_ADRE}

Here, a simple equation with a linear source term is considered,
\begin{linenomath}\begin{align}
\frac{\partial \phi}{\partial t} + \nabla\cdot(\bs{u}\phi)  = \nabla \cdot (M \nabla \phi) + \underbrace{\lambda (\eta(\bs x) - \phi)}_{Q=Q(\phi)}, \label{eq:adr_equation}
\end{align}\end{linenomath}
where $\lambda$ is a constant and $\eta(\bs x)$ is a known function. 
On a periodic domain with constant velocity, $\bs u$, the problem is a first order linear differential equation, 
which acts independently on all wavelengths.
That means that for any selected wavevector $\bs k$, 
we can solve analytically the equation for initial condition $\phi\mid_{t=0} = P e^{\thei \bs k\cdot \bs x}$ 
and $\eta = G e^{\thei \bs k\cdot \bs x}$, where upright $\thei$ denotes the imaginary unit.
The analytical solution will be a transition between the initial condition and the steady state,
\begin{linenomath}\begin{equation}
\phi_\text{analytical}(\bs x, t)=\left(e^{-at}P + (1-e^{-at})a^{-1}\lambda G\right)e^{\thei \bs k\cdot\bs x},\label{eq:adr_solution}
\end{equation}\end{linenomath}
where $a = \lambda + \thei \left(\bs{u}\cdot\bs k\right) + M\left(\bs k\cdot\bs k\right)$.
As the equation is linear, one can take the imaginary or real part of the above analytical solution to obtain a real valued solution.
If not specifically mentioned, the real part is used in tests.
By varying $P$ and $G$ one can study the influence of the initial condition and the steady-state solution respectively.

\subsubsection{Second-order convergence}
\begin{figure}[htbp]
    \centering
    \includegraphics[scale=1]{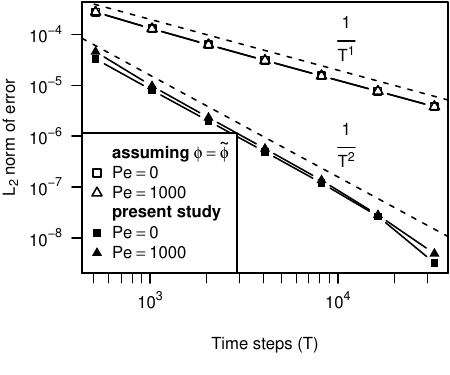}
    \caption{Convergence of the solution of~\cref{eq:adr_equation} compared with analytical solution 
    given by~\cref{eq:adr_solution}, for $k=1\frac{2\pi}{L}$, $P=1$ and $G=0$.}
     \label{fig:linear_decay_1}
\end{figure}

To test the convergence, the~\cref{eq:adr_equation} was solved on a periodic domain of size $L\times L$ elements for a time $t$ with $\mathbf{Fo}=0.001$ and $\mathbf{Da}=1000$.
In all computations of this case, the \ac{SRT} collision operator was used.
The number of time steps was increased by a factor of 2 from $T=2^{9}$ to $T=2^{15}$, while maintaining the number of elements per length, $L$, proportional such that $16L=T$ (acoustic scaling).
The velocity was varied between $\mathbf{Pe}=0$ and $\mathbf{Pe}=1000$ and the wave number $k$ was $0\frac{2\pi}{L}$, $1\frac{2\pi}{L}$ or $2\frac{2\pi}{L}$.
For each setup, two cases were executed. One with $P=1$ and $G=0$, and the other with $P=0$ and $G=1$.
The \ac{LBM} solutions were compared with the analytical solution of~\cref{eq:adr_equation}, and the $\mathcal{L}_2$ norm of the difference was computed.
The proposed scheme consistently recovered second-order convergence for all the cases, with the slope varying from~$1.99$ to~$2.2$ as calculated with a least square fit.
Figure~\ref{fig:linear_decay_1} provides an indication of the convergence for $k=1\frac{2\pi}{L}$, $P=1$ and $G=0$.
The convergence observed was compared to results obtained when ignoring the implicit~\cref{eq:tilde_phi_from_phi} and assuming $\phi=\tilde\phi$ in the calculation of $Q$.

\subsubsection{Convergence of \ac{DBE} to \ac{ADRE}}
The \ac{LBM} can be treated as a discretisation of the \ac{DBE}, which in turn converges to the \ac{ADRE} that one originally wanted to solve.
This means that two types of error need to be considered, namely, the error of discretisation, and the error (difference) between the \ac{DBE} and \ac{ADRE}.
Readers interested in a more detailed study regarding equivalent partial differential equations for the lattice Boltzmann schemes are referred to the recent work of
\citet{Fucik2021} and \citet{SSimonis_2023}.

In the case of the linear source term, $\bs q=Q(\phi) \bs{\gamma( \bs{u})}$, $\bs h^\eq=\phi\bs\gamma( \bs{u})$ and $\bs \gamma( \bs u) = \mathbb{M}^{-1}\bs \Gamma(
\bs u)$; see \cref{eq:Gamma} for the definition of $\Gamma(\bs u)$.
Assuming an uniform velocity field, $\bs u$, the \ac{DBE} given in~\cref{eq:DBE} can be solved analytically and its solution can be conveniently expressed using matrix exponents,
\begin{linenomath}\begin{equation}
\bs h_\text{analytical}^\text{DBE}(\bs x, t)=\left(e^{-\bs At}\bs\gamma(\bs u)P + (I-e^{-\bs At})\bs A^{-1}\bs\gamma(\bs u)\lambda G\right)e^{\thei\bs k\cdot\bs x},\label{eq:dbe_solution}
\end{equation}\end{linenomath}
where $A_{jk} = \thei \delta_{jk}\left(\bs e_j\cdot\bs k\right) - \frac{1}{\tau}\left(\gamma_j(\bs u) - \delta_{jk}\right) + \lambda\gamma_j(\bs u)$.
Substitution of all variables indicates that the analytical solution of the \ac{ADRE} is independent of $L$ and $T$, as one would expect, however, the analytical solution of the \ac{DBE} is only dependent on ratio of $L$ to $T$ (the velocity scale). 

\begin{figure}[htbp]
    \centering
    \includegraphics[scale=1]{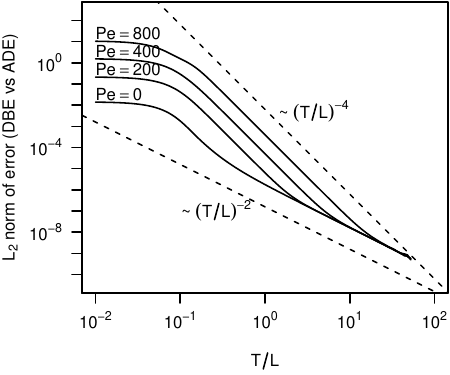}
    \caption{$\mathcal{L}_2$ norm of the difference between the analytical solutions of the discrete Boltzmann~\cref{eq:DBE} and the advection-diffusion-reaction~\cref{eq:ADE}. 
    Evaluated for $k=1\frac{2\pi}{L}$, $M=0.001\frac{L^2}{T}$, $G=0$, and $P=1$.}
     \label{fig:dbe_ade_error}
\end{figure}

Figure~\ref{fig:dbe_ade_error} provides the $\mathcal{L}_2$ norm of the difference between the complex solutions of \ac{DBE} and \ac{ADRE}.
The two components of the error can be observed, namely, the fourth-order, velocity dependent error, and second-order, velocity independent error.

\subsubsection{Error landscape}\label{sec:error_landscape}
A better understanding of the behaviour of the \ac{LBM} method for the advection-diffusion-reaction equation, and \ac{LBM} in general, can be gained by looking at the dependence of the error on both spatial and temporal resolution. 
In this work, this is termed the error landscape.
For most physical problems it is prohibitively expensive to calculate the full error landscape, even if an analytical solutions is available.
Nevertheless, the landscape for this case is discussed here, as the general trends and slopes of this landscape will be similar in any \ac{LBM} application.

\begin{figure}[htbp] 
    \centering
    \includegraphics[scale=1]{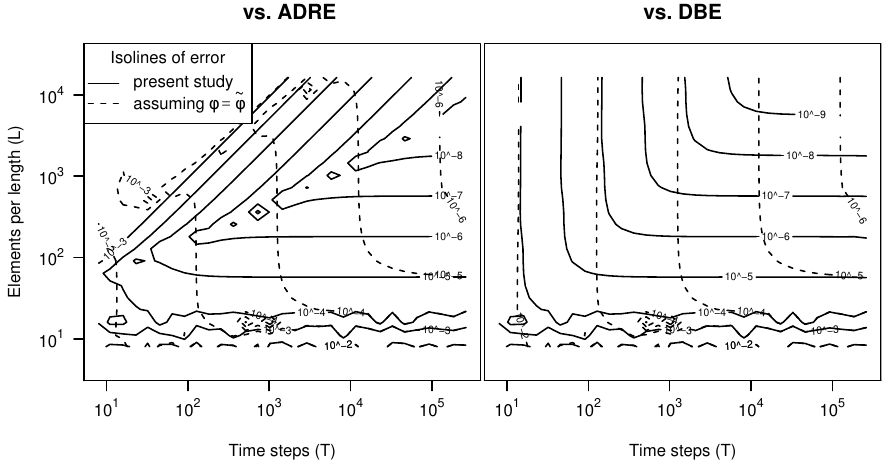}
    \caption{
     The isolines of error of the \ac{LBM} solution against \acl{ADRE} and \acl{DBE}, given by ~\cref{eq:adr_solution} and ~\cref{eq:dbe_solution}, in the T--L space.
     The y-axis of the left-hand and right-hand side plots are the same. 
     Error is defined as $\mathcal{L}_2$-norm of the difference between \ac{LBM} solution and the reference one. 
     All combinations of $23$ and $31$ distinct values of $L$ and $T$ respectively, were simulated. 
     Evaluated for $k=\frac{2\pi}{L}$, $M=0.001\frac{L^2}{T}$, $G=0$, $P=1$, and $u = 0$.}
     \label{fig:error_landscape_real}
\end{figure}

A set of 23 and 31 distinct values of $L$ and $T$ were selected, generating a solution set that is close to a linear distribution in log space.
For each pair, two simulations were performed, one with the method presented in this paper, and one in which $Q$ is calculated from $\tilde\phi$, not $\phi$.
In total, 1426 simulations were performed to populate the error landscape.
The results were compared to both the analytical solutions of the \ac{ADRE} and \ac{DBE}.
Figure~\ref{fig:error_landscape_real} presents the isolines of the error.
One can observe, that the convergence to the \ac{DBE} is only driven by competing temporal and spatial discretisation errors.
On the other hand, the $\nicefrac{T}{L}$ dependent error between \ac{DBE} and \ac{ADRE} dominates the temporal error in convergence to \ac{ADRE}.
In either case, if the source term is inappropriately integrated, its first-order error reduces the accuracy greatly in almost the entire landscape (marked with dashed lines).

\begin{figure}[htbp]
    \centering
    \includegraphics[scale=1]{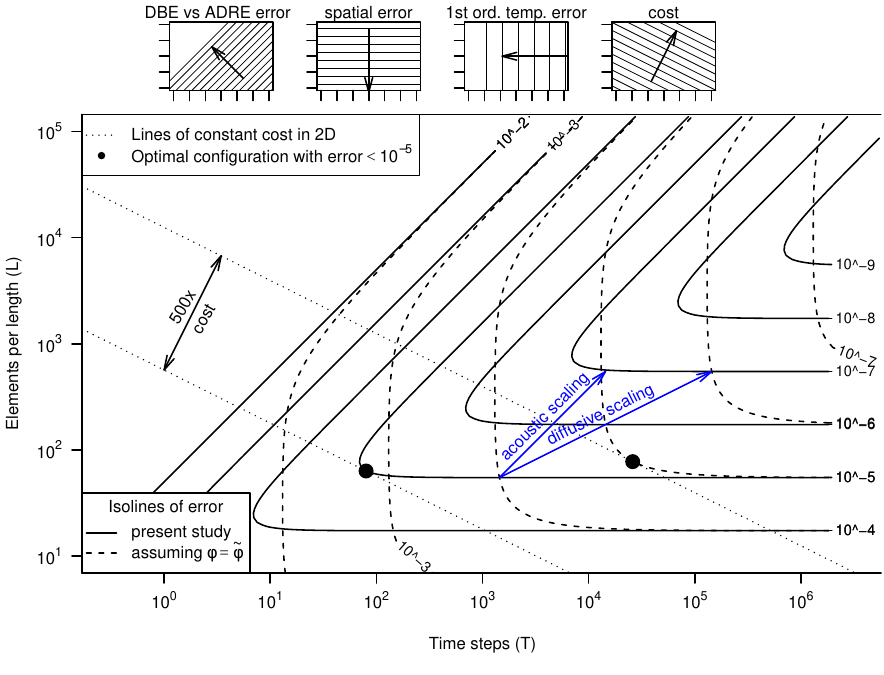}
    \caption{The schematic isolines of error in the T--L space. 
    See Figure~\ref{fig:error_landscape_real} for the plot of actual error, measured in simulation. 
    First three of the top four plots show the trends of: 
    error between \acl{DBE} solution and \acl{ADRE} $\left(\sim\left(\nicefrac{T}{L}\right)^{-2}\right)$, 
    error of the spatial discretisation ($\sim L^{-2}$), 
    1st order temporal error introduced by improper integration of the source term ($\sim T^{-1}$). 
    The last of the top four plots presents the trend of the cost of the simulation for 2D lattice ($\sim L^2T$). 
    The large black circles mark optimal (CPU cheapest) selections of $L$ and $T$ to achieve the error of no more then $10^{-5}$. 
    The ratio of the computational cost for the optimal setup in the naive approach ($\phi=\tilde\phi$) compared to the present study is approximately $500$ times.
    The vectors indicate the direction of spatio-temporal refinement using either acoustic or diffusive scaling.
    Assuming that the length of a vector corresponds to a single refinement step, the second order convergence can be deduced (as two iso-lines of errors are crossed) when looking at the diffusive scaling and setting $\phi=\tilde\phi$.
In other words, the slope of convergence for diffusive scaling would be the same for both the proper and naive implementation.}
     \label{fig:error_landscape}
\end{figure}

In order to provide a clear view of the trends present in the error landscape, approximations of all the separate errors were established to create a smoothed landscape presented in Figure~\ref{fig:error_landscape}.
To quantify the impact of the source term discretisation, the computational cost was compared to achieve the same level of error using either a first- or second-order scheme.
Both for the present method, and for the inconsistent integration, an optimal selection of number of time-steps, $T$, and number of elements, $L$, was made, to achieve an error of no more than $10^{-5}$.
In this specific example, the use of the inconsistent integration scheme for the source term 
leads to an increase in computational effort ($L^2T$) of the factor of 500 to achieve an equivalent level error.

The landscapes presented in~\cref{fig:error_landscape_real,fig:error_landscape} indicate the pitfalls of analysing the convergence of LBM.
It has to be reiterated that the Boltzmann equation is integrated along the characteristics, thus the space and time integration can not be treated independently in the construction of a conventional \ac{LBM} scheme. 
As a consequence, a properly implemented \ac{LBM} scheme has second-order convergence.
The isolines of error, presented in~\cref{fig:error_landscape}, are commonly traversed along the directions marked by the acoustic and diffusive scaling. 
Given a specific scaling, i.e. ratio of temporal to spatial resolution between subsequent refinements, 
a researcher may get a biased view of the error and the order of convergence.
For instance, under the diffusive scaling, a second order convergence in space will be observed for both the trapezoidal and Euler's implementation of the reaction term integrator. 
An example of first order implementation leading to the aforementioned behaviour can be found in \cite{Zhang2019}.
Next, the second order convergence in time is clearly visible in the acoustic scaling, 
while the diffusive scaling works as first order in time (see~\cref{fig:Convergence_study_2D}).
Moreover, in the lower right corner of the landscape, the error caused by Euler's implementation is relatively small, 
thus it may not affect the order of convergence.
Concluding, the same gain in accuracy can be accomplished along different pathways. 
The choice of the pathway will influence the computational cost.
Based on this, one could select the spatio-temporal parameters of the simulation to progress along the iso-error line in order to obtain the desired result with a lower computational cost.

\subsection{The Allen-Cahn equation --- illustrative advection-diffusion-reaction problem.\label{sec:AllenCahn}}
\begin{figure}[htbp]
\centering
   \includegraphics[width=0.5\textwidth]{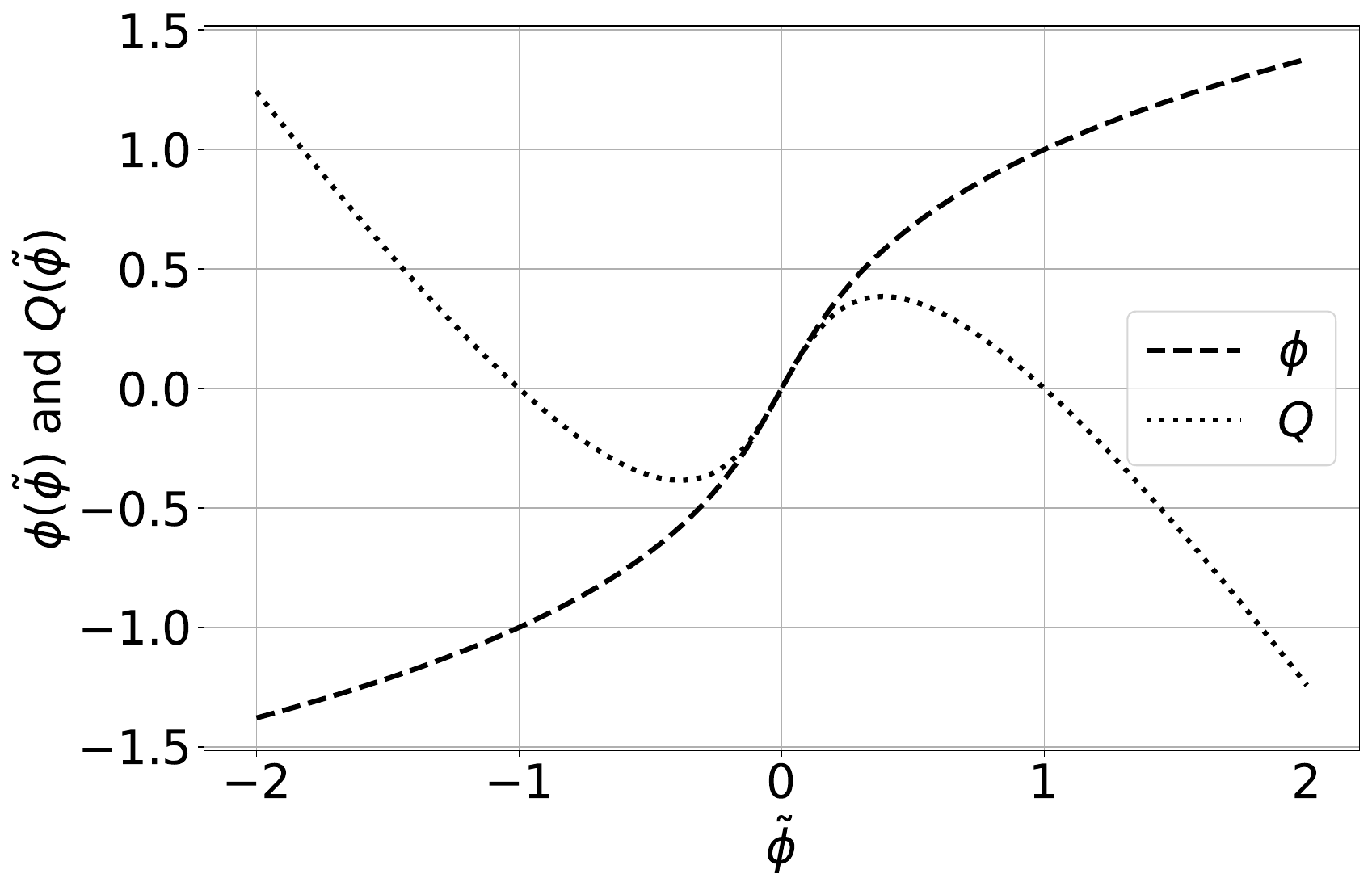} 
   \caption{Plot presents the real solution of \cref{eq:ac_phi}, \\
    namely $\phi=\phi(\tilde{\phi})$ and $ Q=Q(\tilde{\phi})$ for $\lambda=1$.} \label{fig:AC_roots}
\end{figure}

In this section a solidification problem was selected in order to illustrate the impact of the non-linear dependence of the source term on the transported  scalar 
field.
The problem solved is the Allen-Cahn equation in the form,
\begin{linenomath}\begin{align}
\frac{\partial \phi}{\partial t} + \nabla\cdot(\boldsymbol{u}\phi)  = \nabla \cdot (M \nabla \phi) + \underbrace{\lambda \phi \left(1-\phi^{2}\right)}_{Q=Q(\phi)}, \label{eq:Allen-Cahn_equation}
\end{align}\end{linenomath}
where the source term is responsible for the phase change.
The details for this equation can be found in the dedicated literature, for example \citet{cahn1958freeenergy,jacqmin1999calculation}.
To formulate the implicit relation between $\phi$ and $\tilde\phi$, the source term, $Q(\phi)$, was substituted into \cref{eq:tilde_phi_from_phi},
\begin{linenomath}\begin{align}
\tilde{\phi} &= \sum_i \lbmh_i = \phi - \dfrac{1}{2} Q(\phi) = \phi - \dfrac{1}{2}\lambda \phi \left(1-\phi^2\right) 
= \phi \left(1 - \dfrac{\lambda}{2} \left(1-\phi^2 \right)\right). \label{eq:binding_phi_and_tilde_phi}
\end{align}\end{linenomath}
As previously discussed, this equation has to be solved to express $\phi$ as a function of $\tilde\phi=\sum_i h_i$ to resolve the implicit relation introduced by $Q = Q(\phi)$.
The exact solution can be readily derived for this third order polynomial.
For $\lambda<2$, there is only one real-valued solution,
\begin{linenomath}\begin{equation}
\phi(\tilde{\phi}) = \frac{A}{C} - C
\text{, where }A=\frac{2-\lambda}{3\lambda}
\text{, }B = \frac{\tilde\phi}{\lambda}
\text{, and }C = \sqrt[3]{\sqrt{B^2 + A^3} - B}.\label{eq:ac_phi}
\end{equation}\end{linenomath}
In the next sections, this analytical expression is used in the \ac{LBM} collision operator to calculate $\phi$, and $Q(\phi)$.
~\Cref{fig:AC_roots} illustrates the relationship between $\phi$, $Q$ and $\tilde{\phi}$.
Two stable fixed points correspond to the roots of $Q$, present at $\phi=\tilde\phi = 1$ and $-1$, and one unstable at $0$. 
The round-off errors can cause the solution of~\cref{eq:binding_phi_and_tilde_phi} to drift from the unstable, 0, root (see~\cref{fig:AC_roots}).
Although direct implementation (as in the present work) of~\cref{eq:ac_phi} and calculation of the source term as $Q=Q(\phi)=\lambda \phi \left(1-\phi^{2}\right)$ generates acceptable results, one can seek to improve the stability in two ways.
In the first approach, an iterative (e.g. Newton-Raphson) method can be applied to find $\phi$ directly from~\cref{eq:binding_phi_and_tilde_phi} instead of~\cref{eq:ac_phi}. 
Then, $Q=Q(\phi)$ is calculated as before.
Alternatively, $\phi$ and $\tilde{\phi}$ can be calculated from~\cref{eq:ac_phi} and~\cref{eq:binding_phi_and_tilde_phi} respectively, 
while the source term (see~\cref{eq:tilde_phi_from_phi}) would correspond to double of their difference, $Q=2(\phi - \tilde{\phi})$.

The following subsections are ordered by the growing complexity of benchmarks.
First, only the reaction term is benchmarked.
Then, a non-uniform initial condition is applied to observe the diffusive effects.
Finally, an external velocity field is imposed to obtain the full advection-diffusion-reaction problem.

\subsubsection{Uniform reaction benchmark --- comparison with an analytical solution}\label{sec:AllenCahn_uniform_IC}
This section analyses the evolution of a uniform initial distribution of the scalar field, $\phi$, in the absence of an external velocity field. 
The spatial derivatives in \cref{eq:Allen-Cahn_equation} reduce to zero and the problem simplifies to an ordinary differential equation,
\begin{linenomath}\begin{align}
\frac{d\phi}{dt} = \lambda \phi(1-\phi^2).\label{eq:AC_uniform_IC}
\end{align}\end{linenomath}
The analytical solution of this equation is,
\begin{linenomath}\begin{align}
\phi{\left(t \right)} = \pm \left(C_{1} e^{- 2 \lambda t} + 1\right)^{-\frac{1}{2}} & \text{, where } C_{1} = \left(\phi(0)\right)^{-2} - 1.
\end{align}\end{linenomath} 

\begin{figure}[htbp]
    \centering
    \begin{subfigure}[c]{0.475\linewidth}
        \centering
        \includegraphics[width=1\linewidth]{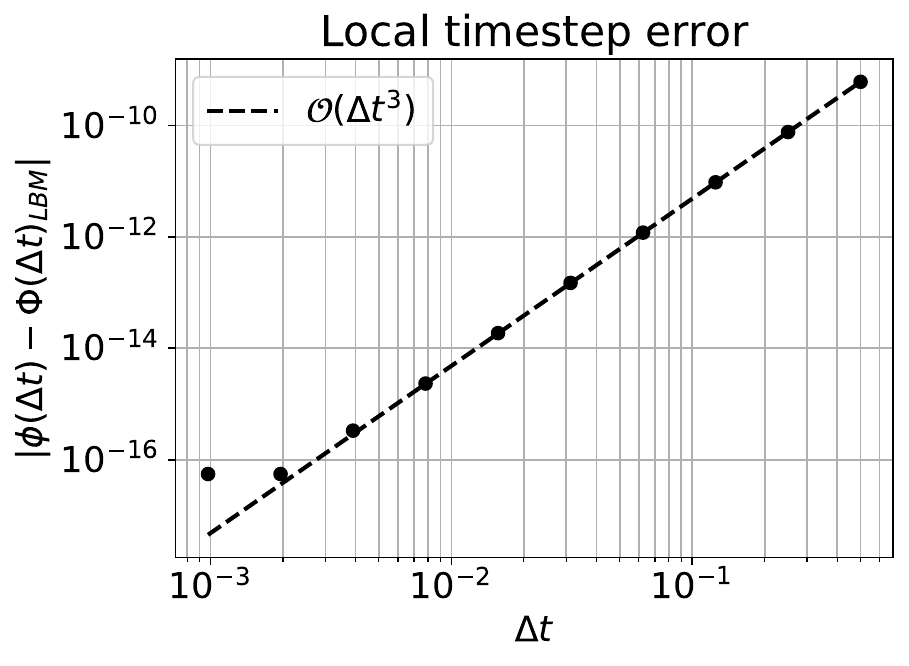}
        \caption{Local truncation error for a single time step. The error reaches $10^{-16}$ being the limit of numerical accuracy.}
	\label{fig:conv-0d-local}
    \end{subfigure}
    ~ 
    \hspace{1em}
    \begin{subfigure}[c]{0.475\linewidth}
        \centering
        \includegraphics[width=1\linewidth]{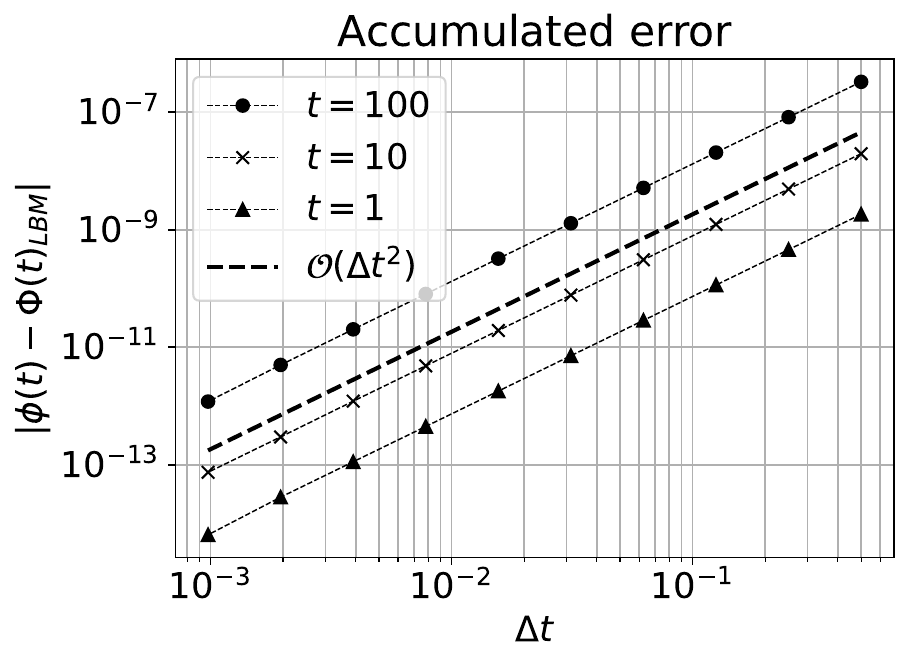}
        \caption{Convergence of total/accumulated error, evaluated for three different times $t=1;10;100$.}
        \label{fig:conv-0d-global}
    \end{subfigure} 
    \caption{Convergence study of the reaction component of the Allen-Cahn \cref{eq:Allen-Cahn_equation}, 
    on a $D2Q9$ lattice with uniform initial condition and on periodic domain with $\lambda=0.01$. 
    In such a case, the problem simplifies to an ODE given by \cref{eq:AC_uniform_IC}. 
    The operator $\left|\bullet \right|$ on the vertical axis denotes a scalar absolute value.
    It is used to evaluate the error against the analytical solution.
    In general, a method which converges with $\mathcal{O}(n+1)$ local truncation error, has a global error of order $\mathcal{O}(n)$. 
    In this example, the convergence of a local time-step error is limited by truncation error at $\approx 10^{-16}$.
    \label{fig:conv-0d}}
\end{figure}

The results obtained from \ac{LBM} using the~\cref{eq:ac_phi}, were compared with the analytical solution of the \ac{ODE}.
In~\cref{fig:conv-0d}, both local and global (accumulated) truncation errors are presented.
The local time-step error was calculated as the difference between the numerical and analytical solution after a single time step. 
For this case, the third-order convergence up to the level of computational accuracy was recovered.  
As there is no diffusion, the \ac{Fo} is undefined. 
The global convergence rate was determined by comparison of the numerical and analytical solution after a fixed time $t \in {1, 10, 100}$. 
The coefficient responsible for the intensity of reaction was set to $\lambda=0.01$.

As observed in~\cref{fig:conv-0d-global}, the scheme recovered the expected second-order convergence. 
From this result it can be concluded that the presented implementation recovers the trapezoidal integration scheme for the special case in which the Allen-Cahn \ac{PDE} reduces to an \ac{ODE}.

It is reminded that to obtain accurate solution, an accurate initialisation procedure which took into account the \textit{shift} in calculated values (see~\cref{eq:initialization}) must be applied.

\subsubsection{Reaction-diffusion benchmark (2D, periodic) --- comparison with a finite-element solution}\label{sec:fem_lbm}
\begin{figure}[htbp]
    \centering
    \begin{subfigure}[c]{0.5\linewidth}
        \centering
        \includegraphics[width=1\linewidth]{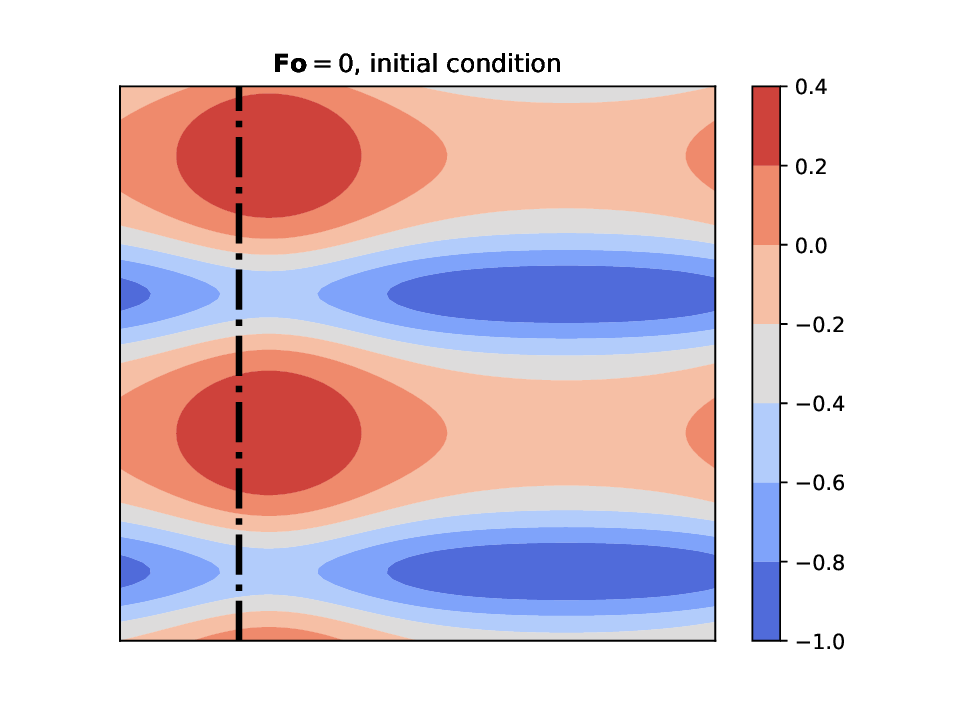}
        \caption{The initial condition, given by \cref{eq:initial_condition}, used to initialize the intensity of the phase field, $\phi$, for both the \ac{FEM} and \ac{LBM} solvers.\label{fig:initial_img}}
    \end{subfigure}%
    ~
    \hspace{1em}
    \begin{subfigure}[c]{0.5\linewidth}
        \centering
        \includegraphics[width=1\linewidth]{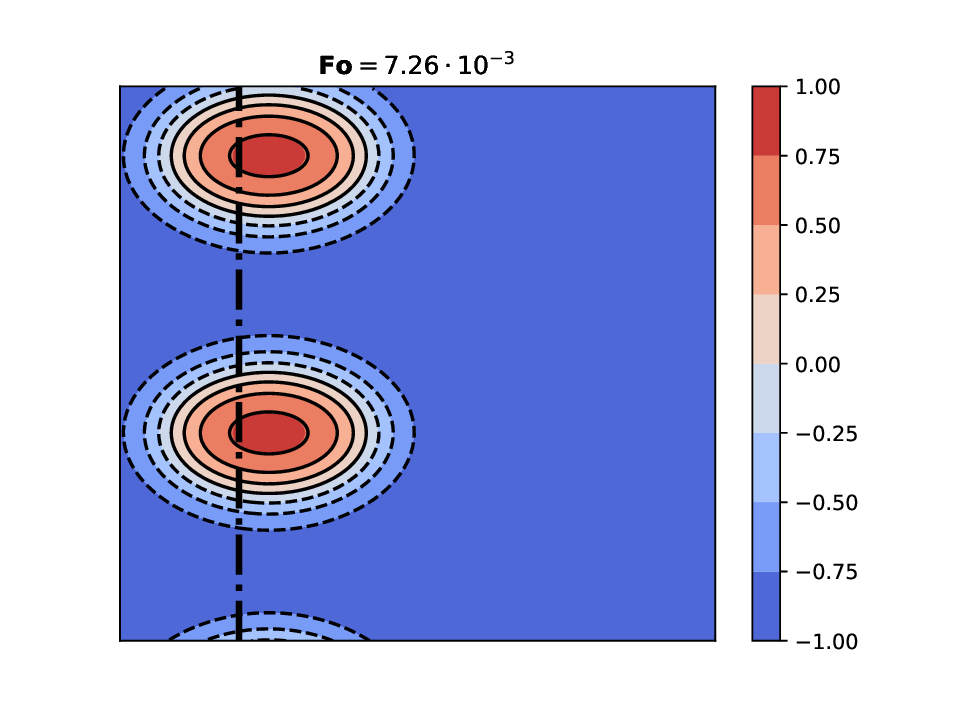}
        \caption{The intensity of the phase field, $\phi$, at the final time, $\text{\acs{Fo}}= \num{7.26 e-03}$.
         Both \ac{FEM} (colors) and \ac{LBM} (lines) solutions are presented. \label{fig:fem_lbm_cmp_field}}
    \end{subfigure}
    \vspace{1em}
    \\
    \begin{subfigure}[c]{0.5\linewidth}
        \centering
        \includegraphics[width=1\linewidth]{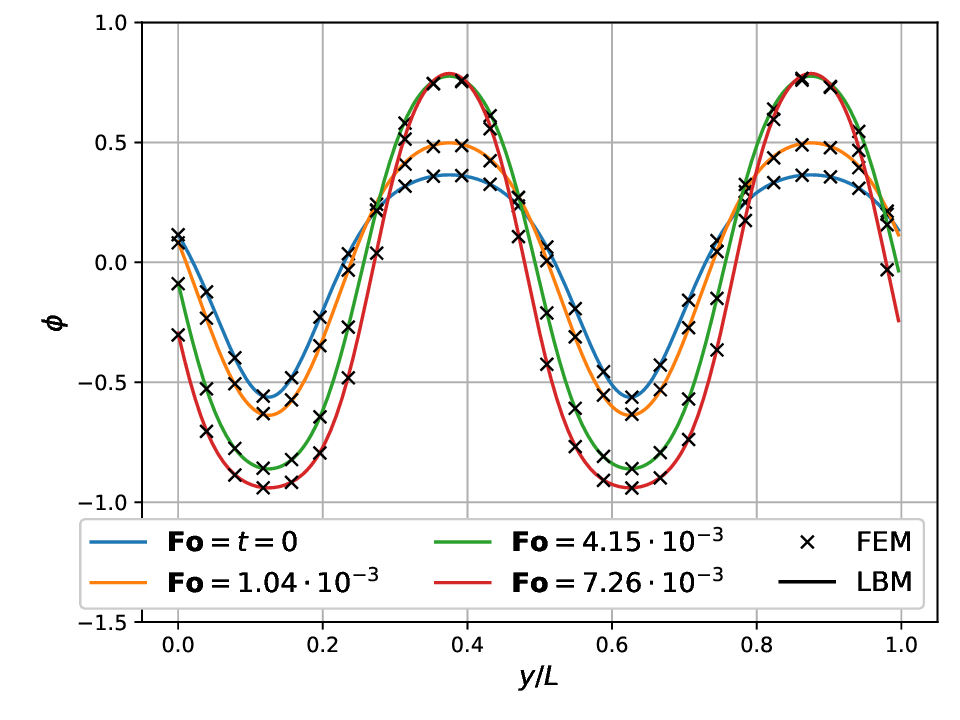}
        \caption{The time evolution of the phase field, $\phi$, 
        is captured for four different Fourier numbers at a cross-section defined at $x/L = 0.2$.
        The \ac{FEM} and \ac{LBM} solutions are marked with crosses and lines respectively.
        }
    \label{fig:fem_lbm_cmp}
    \end{subfigure}
    \caption{Comparison of the \ac{FEM} and \ac{LBM} solutions for reaction-diffusion problem specified in section \ref{sec:fem_lbm}. 
    The \ac{LBM} lattice consists of $256x256$ nodes.
    Domain is periodic.
    The P{\'e}clet and Damk{\"o}hler numbers are  $\text{\acs{Pe}}=0, \text{\acs{Da}}= 500$.
    }
\end{figure}

To ensure that the proposed collision kernel properly recovers the diffusion process, 
the \ac{LBM} solver has been qualitatively compared against the \acf{FEM}.
A spatially varying, periodic initial condition was applied on a square, unit domain using the exponential function, 
\begin{linenomath}\begin{equation}
\phi\mid_{t=0}(x,y) = \frac{1}{2e-e^{-1}}\left(e^{\sin{\left( 2\pi \frac{x}{L} \right)}} -  2 e^{\sin{ \left( 4\pi \frac{y}{L} \right) }}\right). \label{eq:initial_condition}
\end{equation}\end{linenomath}
\cref{fig:initial_img} shows the result of the initialisation.

The \ac{FEM} results were obtained using the FEniCS~\cite{AlnaesBlechta2015a} solver. 
We used fourth-order Lagrange interpolation for spatial discretisation and fourth-order Runge-Kutta (ESDIRK43a) with adaptive time-stepping for time integration.
The \ac{FEM} was solved on a regular, triangular mesh in the 2D square domain of unit length. There were 25 divisions along each side of the domain resulting in $\sim 26^2$ elements in total.
The time step ranged from approximately $10^{-4}$ to $10^{-5}$.
Diffusivity in \ac{FEM} solver was set to one, while the reaction rate was defined by setting $\text{\acs{Da}}=500$. 
The \ac{LBM} domain was discretised with $256 \times 256$ lattice nodes and the diffusivity was set to $1/6$ in lattice units, and magic number was set to $1/12$.
To match the solutions obtained by two different solvers, the results were reported for a specific non-dimensional time.
There was no external velocity field, thus $\text{\acs{Pe}}=0$.

\Cref{fig:fem_lbm_cmp_field} provides a qualitative comparison between the \ac{FEM} and \ac{LBM} solutions at a time defined by $\text{\acs{Fo}} = \num{7.26e-03}$.
Here, the colour contours from the \ac{FEM} solution are aligned with the dashed iso-lines of the \ac{LBM} solution.
\Cref{fig:fem_lbm_cmp} quantitatively compares the time evolution of the scalar field, $\phi$, computed by both solvers.
through the cross section denoted by the vertical dashed line in~\cref{fig:initial_img} and~\cref{fig:fem_lbm_cmp_field}. 
The initial condition, and three subsequent times are provided in the same figure.
It can be clearly seen, that the \ac{LBM} and \ac{FEM} solutions agree.

\subsubsection{Advection-diffusion-reaction benchmark (2D, periodic) ---  self-convergence study} \label{sec:periodic_2D_Da_convergence}
This section investigates the convergence of the advection-diffusion-reaction problem, given by~\cref{eq:Allen-Cahn_equation}.
The exponential initial condition was used again, as per \cref{eq:initial_condition}.
For this study, a \ac{TRT} collision operator was used again, with the magic coefficient, $\Lambda$, 
set to $1/12$ in order to cancel out the third-order spatial error and provide optimal results for advection dominated problems \cite{Ginzburg2008consistentLBM4BrinkmanModel}.
The scalar field was advected with a uniform external velocity, $U_x$, corresponding to $\text{\acs{Pe}} = \num{1e03}$. 
The time of the simulation was determined by setting $\text{\acs{Fo}} = \num{1e-06}$.
The \acl{Da} has been set as $\text{\acs{Da}} = \num{1e06}$, to provide a scenario in which the reaction term dominates the diffusion behaviour.
The convergence of the scheme was assessed using both acoustic and diffusive scaling techniques. 
The parameters used during these studies are provided in \cref{Tab:parameters_acoustic_scaling} and \cref{Tab:parameters_difusive_scaling}, respectively. 
Here, the number of elements per the domain length, $L$, is provided, as is the number of iterations, $T$. 

\begin{table}[htbp]
\centering
\caption{Simulation parameters in lattice units, acoustic scaling, for $\mathbf{Pe}=\num{1e03}$, $\mathbf{Fo}=\num{1e-06}$ and $\text{\acs{Da}}=\num{1e06}$.}
\begin{tabular}{rr|rrrrr}
\toprule
 L & T & $U_x$ & M & $\lambda $ \\
\midrule
4096 &          1024 & \num{4.00e-03} & \num{1.64e-02} &  \num{9.77e-04} \\
2048 &           512 & \num{4.00e-03} & \num{8.19e-03} &  \num{1.95e-03} \\
1024 &           256 & \num{4.00e-03} & \num{4.10e-03} &  \num{3.91e-03} \\
 512 &           128 & \num{4.00e-03} & \num{2.05e-03} &  \num{7.81e-03} \\
 256 &            64 & \num{4.00e-03} & \num{1.02e-03} &  \num{1.56e-02} \\
\bottomrule
\end{tabular}
\label{Tab:parameters_acoustic_scaling}
\end{table}

\begin{table}[htpb]
\centering
\caption{Simulation parameters in lattice units, diffusive scaling, for $\mathbf{Pe}=\num{1e03}$, $\mathbf{Fo}=\num{1e-06}$ and $\text{\acs{Da}}=\num{1e06}$.}
\begin{tabular}{rr|rrrrr}
\toprule
 L & T & $U_x$ & M & $\lambda$  \\
\midrule
4096 &          1024 & \num{4.00e-03} & \num{1.64e-02} &  \num{9.77e-04} \\
2048 &           256 & \num{8.00e-03} & \num{1.64e-02} &  \num{3.91e-03} \\
1024 &            64 & \num{1.60e-02} & \num{1.64e-02} &  \num{1.56e-02} \\
 512 &            16 & \num{3.20e-02} & \num{1.64e-02} &  \num{6.25e-02} \\
 256 &             4 & \num{6.40e-02} & \num{1.64e-02} &  \num{2.50e-01} \\
\bottomrule
\end{tabular}
\label{Tab:parameters_difusive_scaling}
\end{table}

\begin{figure}[ht]
    \centering
    \begin{subfigure}[b]{0.49\linewidth}
        \centering
        \includegraphics[width=1\linewidth]{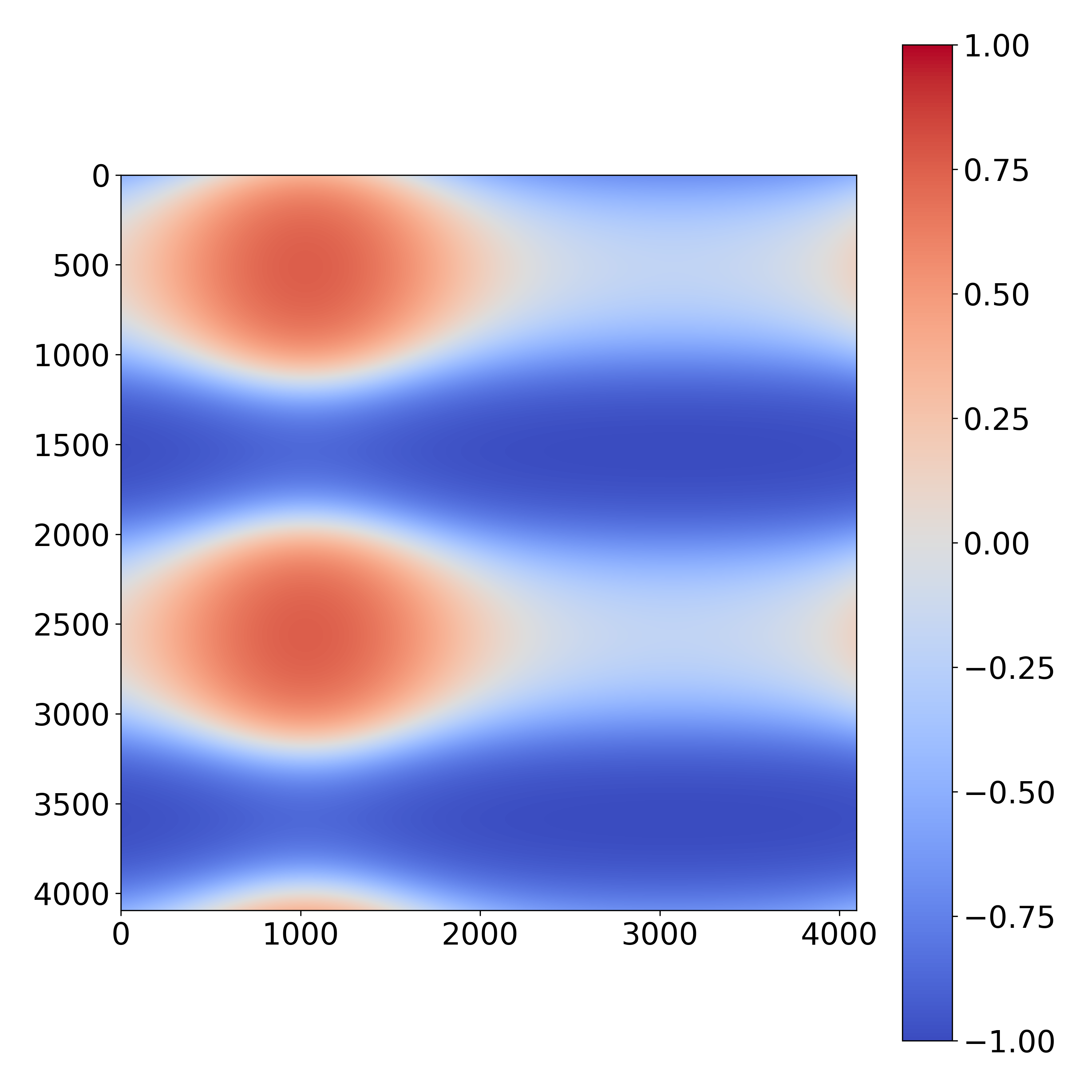}
        \caption{Phase field, $\phi$.}
    \end{subfigure}
    ~ 
    \begin{subfigure}[b]{0.49\linewidth}
        \centering
        \includegraphics[width=1\linewidth]{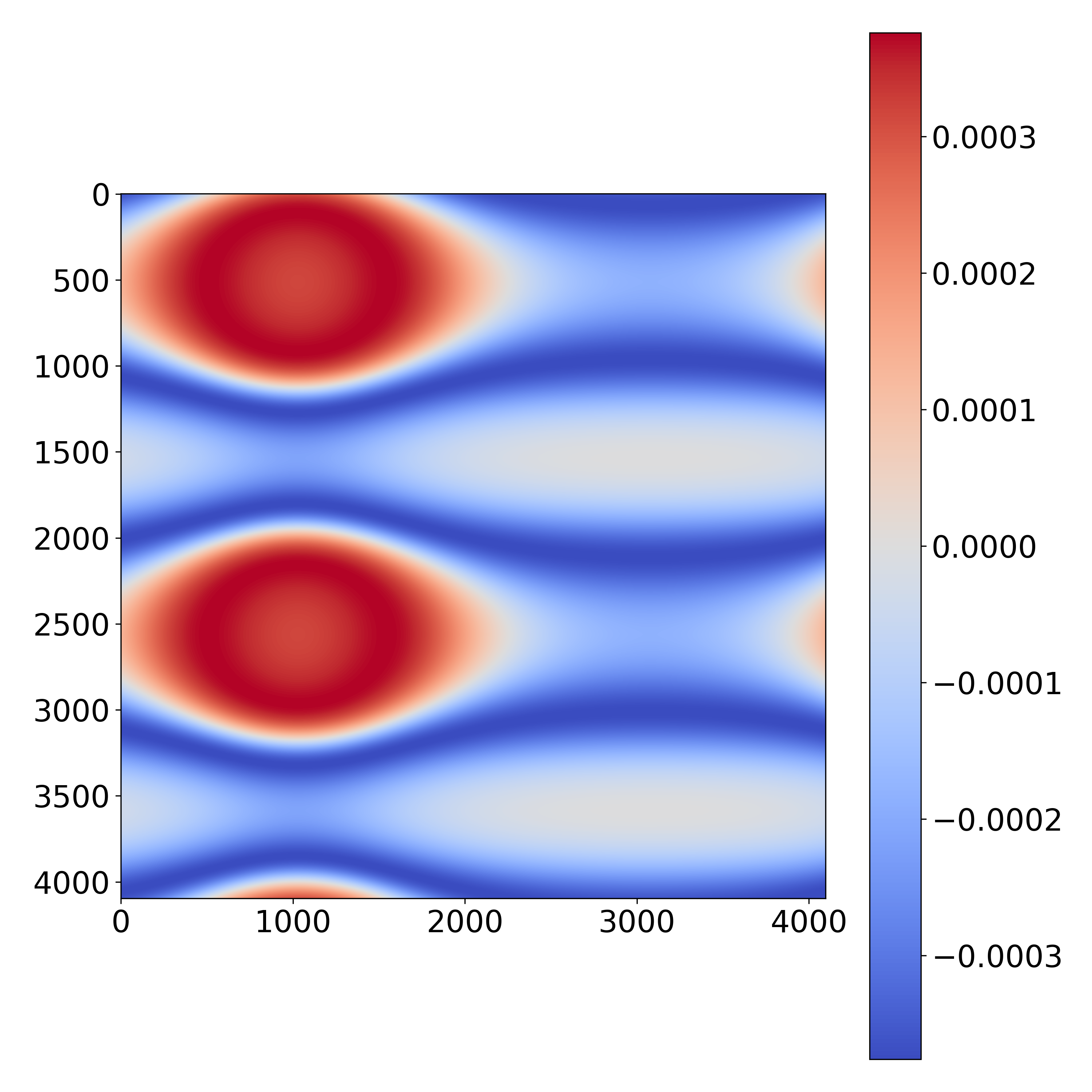}
        \caption{Source term, $Q$.}
    \end{subfigure}
    \caption{Contours of the reference solution on the $4096 \times 4096$ lattice, for $\mathbf{Pe}=\num{1e03}$, $\mathbf{Fo}=\num{1e-06}$ and $\text{\acs{Da}}=\num{1e06}$.
    The solver has been initialized according to~\cref{eq:initial_condition}.
    }
    \label{fig:reference_solution_acoustic_scaling}
\end{figure}

\begin{figure}[ht]
    \centering
    \begin{subfigure}[b]{\textwidth}
        \includegraphics[width=\linewidth]{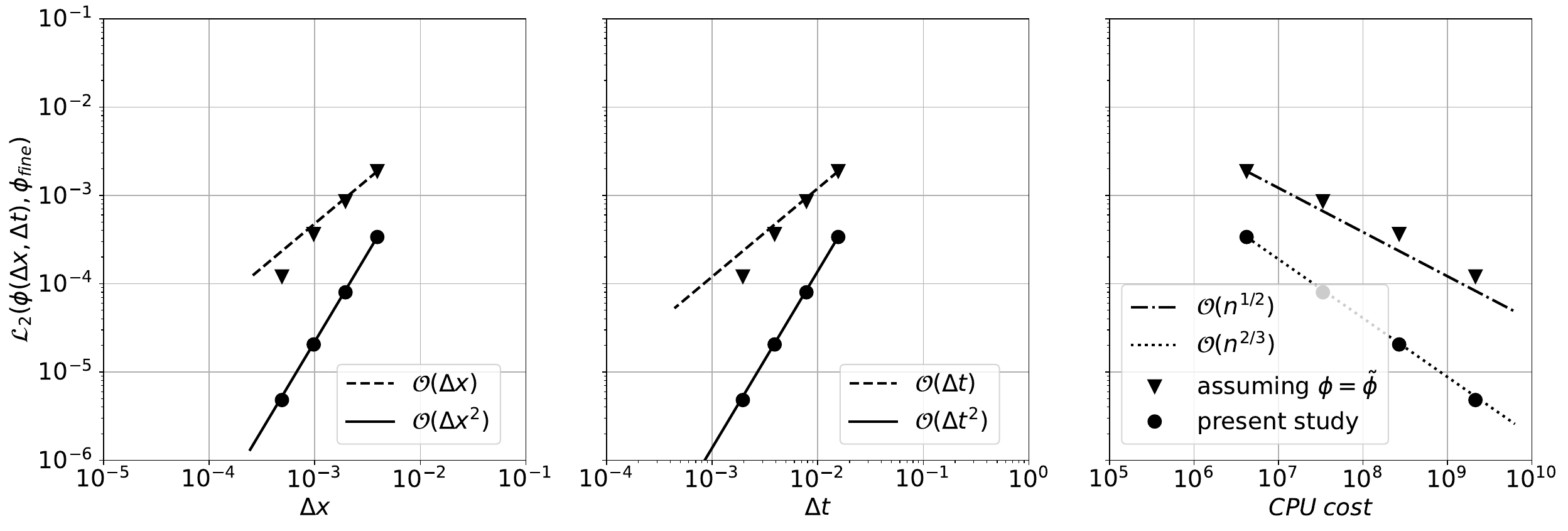}
        \caption{Acoustic scaling}
        \label{fig:Acou_Da_1o00e+06_Pe_1o00e-06}
    \end{subfigure}
    \vskip\baselineskip
    \begin{subfigure}[b]{\textwidth}
		\includegraphics[width=\linewidth]{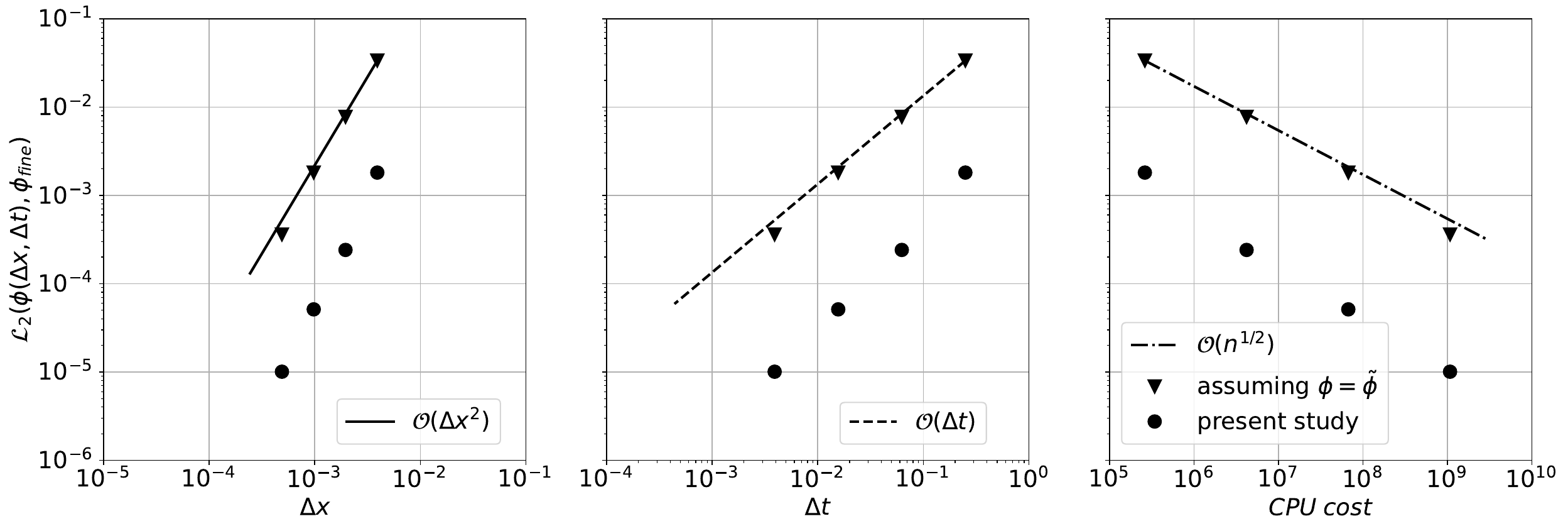}
        \caption{Diffusive scaling}
        \label{fig:Diff_Da_1o00e+06_Pe_1o00e-06}
    \end{subfigure}
    \caption{Self-convergence study of an advection-diffusion-reaction \cref{eq:Allen-Cahn_equation}, on a $D2Q9$ lattice for $\mathbf{Pe}=\num{1e03}$, $\mathbf{Fo}=\num{1e-06}$ and $\text{\acs{Da}}=\num{1e06}$.  
    Notice, that the formally second-order \ac{LBM} scheme works as first-order with respect to time if the refinement is done along the diffusive pathway (see~\cref{sec:error_landscape}).
    Despite the same order of convergence under the diffusive scaling for both proper and naive (assuming $\phi = \tilde{\phi}$) discretization of the source term, the level of $\mathcal{L}_2$ error differs.
}
 \label{fig:Convergence_study_2D}
\end{figure}

\Cref{fig:reference_solution_acoustic_scaling} provides the result of the finest resolution simulated for the scalar variable, $\phi$, and the implicit source term, $Q(\phi)$. 
As there does not exist an analytical solution to the complete advection-diffusion-reaction equation, this finest mesh solution is used to provide a reference point of convergence tests.
The two scaling approaches (acoustic and diffusive) are evaluated in \cref{fig:Convergence_study_2D} for both the proper and naive (assuming $\phi = \tilde{\phi}$) discretization of the source term, $Q$.
The spatial and temporal resolutions are calculated as $\Delta x = 1/L$ and $\Delta t = 1/T$, respectively.
In both scenarios, the proposed scheme provided second-order convergence with respect to the grid spacing.
However, when the diffusive scaling is applied, the spatial discretisation error dominates (see Fig.~\ref{fig:error_landscape}) 
which manifests as first-order dependence of the error on the time step. 
To understand the impact these convergence rates have in practice, the $\mathcal{L}_2$ error norm is shown as a function of the computational cost, defined as $CPU_{cost} = TL^2$, in \cref{fig:Convergence_study_2D}.
Assuming that the simulation parameters will be maintained within the stability regime of the \ac{LBM} (namely, $U_{max} \ll 0.1$, $\omega < 2$), 
one may conclude that for a given set of dimensionless numbers, it is more computationally effective to use acoustic scaling for mesh refinement, and diffusive scaling for mesh rarefaction.
It is noted here, however, that a domain-specific simulation may affect the scaling approach. 
For example, one may prefer to use diffusive scaling in the simulation of flow through a porous medium as the effective boundary location is known to be affected by the choice of relaxation frequency \cite{Pan2006,Regulski2015}.

In addition to the above, the convergence rates reported in this test case highlight the importance of the error landscape previously introduced. 
The rates observed in \cref{fig:Convergence_study_2D}, indicate that the set of simulations were conducted below the diagonal of \cref{fig:error_landscape}.
This apparent diagonal line sets the limit on the accuracy that could be obtained by means of a single acoustically scaled set of grids, and should be noted by practitioners looking to study the convergence of their \ac{LBM} scheme.
For the particular scenario investigated here, the temporal error dominates.
By refining solely the time step, one can hide the source term discretization error below the spatial one and regain the second order convergence under acoustic scaling for the naive scheme.
\clearpage
\acresetall 
\section{Conclusions}\label{sec:conclusions}
Since its initial formulation, the \ac{LBM} has been developed and adapted to solve a variety of equations, addressing different physical scenarios.
Current study has been focused on the details of the usage of \ac{LBM} for \ac{ADRE} in which the source term directly depends on the advected field.
Such source terms arise in the study of heat and mass transfer, phase-transition or evolution of species populations.
Moreover most, implementations of immersed boundary method for heat flows, use source terms to account for desired thermal boundary conditions.
In this paper, the state-of-the-art applications of the \ac{LBM} for \ac{ADRE} were presented, and the main differences in discretisation were discussed.
A clear framework for derivation of the \ac{LBM} numerical scheme from the \ac{DBE} was discussed, and the algebraic manipulations needed for it, were explicitly shown. 
Next, a simplification of the collision operator in the moments' space has been proposed. 

The article reminds that an implicit relation (see \cref{eq:tilde_phi_from_phi}) between the value of the macroscopic field and the zeroth-moment (sum) of the \ac{LBM} densities is the key component for recovering the second-order convergence of the resulting \ac{LBM} numerical scheme.
Furthermore, the closed form solutions of this relation were presented for a variety of common source terms, ranging from simple linear terms to Gompertz model and the Allen-Cahn equation.
Using this implicit relation, the paper presented a local and explicit single- or \ac{TRT} \ac{LBM} for \ac{ADRE} with source terms dependent of the transported field.

To demonstrate the order of convergence of the proposed approach, two distinct equations, involving a linear source term, and a third-order one (as in the Allen-Cahn equation) were investigated.
In the case of the linear source term, analytical solutions for both \ac{ADRE} and \ac{DBE} were obtained, allowing for the study of the dependence of the error of the \ac{LBM} on both spatial and temporal resolution.
This dependence was visualized as isolines, forming an error landscape.
Different competing sources of error were discussed, as was the dependence of the convergence graph on choice of scaling and initial parameters.
Moreover, by investigating the isolines of error one can find the point which minimizes the computational effort for a given accuracy.
In the second benchmark, the relation between the macroscopic field and the shifted one was non-linear and implicit.
The analytical solution was derived and embedded into the explicit, local evaluation of the \ac{LBM} collision operator. 
The resulting numerical scheme was evaluated on three different test problems. 
Firstly, the domain was initialised with a uniform initial distribution to verify the accuracy of the time integration of the source term independently.
The uniform initial distribution removed the spatial derivatives from the governing equation and allowed an analytical solution to be derived.
In this scenario, the proposed scheme recovered the expected second-order global, and third-order local-in-time convergence.
Next, the diffusion term was included in the assessment along with a periodic initial condition based on the exponential function.
The result of a two-dimensional \ac{LBM} simulation was compared to a fourth-order \ac{FEM} solution, as an analytical solution for such case was no longer obtainable.
Finally, to introduce advective effects an external velocity field was added to drive the phase field and a self-convergence study has been performed. 
The \ac{TRT} collision operator was required to reduce the influence of numerical artefacts in the results.

All of the examples confirmed the consistency of the proposed approach and have shown the expected order of convergence.
Observe, that the same implementation of the \ac{LBM} scheme exhibits different order of convergence depending on the ratio of temporal and spatial refinements.
For example, the formally second-order \ac{LBM} scheme displays a first-order convergence with respect to time if the analysed with diffusive scaling (see \cref{fig:Convergence_study_2D}).
Interestingly, in the case of this scaling, both the proper and naive (assuming $\phi=\tilde\phi$) discretizations will exhibit the same {\it slope} of convergence (see \cref{fig:error_landscape,fig:Convergence_study_2D}), but the second-order scheme will result in the lower overall level of error. 
It is thus concluded that great care should be taken when verifying the order of the \ac{LBM} scheme with tests of convergence.

\section{Future Outlook}
The same implicit relation of the macroscopic field to zeroth moment of the \ac{LBM} densities, can be applied in the case of a system of \ac{ADRE}, simply by replacing $\phi$ in \cref{eq:tilde_phi_from_phi} by a vector of fields of interest.
However, detailed analysis of the behaviour of the resulting \ac{LBM} schemes is beyond the scope of this paper.

\section{Declaration of Competing Interest}
The authors declare that they have no known competing financial interests or personal relationships that could have appeared to influence the work reported in this paper.

\section{CRediT authorship contribution statement}
G. Gruszczy\'nski: --- Writing original draft, Review \& Editing, Conceptualization, Methodology, Software, Data curation, Validation, Visualization,
Formal analysis, Investigation.

M. Dzikowski --- Formal analysis, Conceptualization, Investigation.
\ref{app:ODEexample}, \cref{sec:AllenCahn_uniform_IC,sec:fem_lbm} Writing, Software, Data curation, Validation, Visualization.

\L{}. \L{}aniewski-Wo\l{}\l{}k --- Formal analysis, Investigation.
\cref{tab:phi_solutions} and \cref{sec:linear_ADRE}: Writing, Software, Data curation, Validation, Visualization.
			
\section{Acknowledgements}
Authors would like to acknowledge Travis Mitchell for proofreading of the manuscript and Jacek Szumbarski for insightful discussions.
Numerical experiments were performed using computational resources provided by the Interdisciplinary Center for Mathematical and Computational Modelling of the University of Warsaw under grants GR80-12 and GR83-20.
Work was supported by IDUB grant \textit{Modelling of epidemic spreading with Lattice Boltzmann Method}, funded by Warsaw University of Technology.
The simulations were completed using the open-source TCLB solver \cite{Laniewski-Wollk2016a} 
available at: \url{https://github.com/CFD-GO/TCLB}

\appendix
\section{Example: Direct derivation of the 0D scheme from ODE}\label{app:ODEexample}\newcommand{\dt}{\delta t}
As an illustrative exercise, a direct derivation of numerical scheme for an \ac{ODE}, analogous to the proposed \ac{LBM} scheme, is presented. 
Consider an \ac{ODE},
\begin{linenomath}\begin{align}
\frac{d}{dt}\phi=Q(\phi,t).
\end{align}\end{linenomath}
Integrating it with the trapezoidal rule leads to the update formula,
\begin{linenomath}\begin{align}
\phi(t+\dt)=\phi(t)+\frac{\dt}{2}  \bigg( Q(\phi(t),t)+Q(\phi(t+\dt),t+\dt)  \bigg),
\end{align}\end{linenomath}
which belongs to a wider class of implicit Adams-Moulton methods.
Introduction of a $\frac{\dt}{2}$ shift, similarly to the \ac{LBM}
derivation presented in \cref{sec:DBE},
\begin{linenomath}\begin{align}
\tilde{\phi}=\phi-\frac{\dt}{2}Q \longrightarrow \phi = \tilde{\phi} + \frac{\dt}{2}Q,
\end{align}\end{linenomath}
allows to transform this update formula into an explicit one,
\begin{linenomath}\begin{align}
\tilde{\phi}(t+\dt)+\cancel{\frac{\dt}{2}Q(\phi(t+\dt),t+\dt)}=\tilde{\phi}(t)+\frac{\dt}{2}Q(\phi(t),t)+\frac{\dt}{2}\left(Q(t)+\cancel{Q(\phi(t+\dt),t+\dt)}\right),
\end{align}\end{linenomath}
as long as the transformation between $\phi$ and $\tilde\phi$ is bijective.
This lead to a numerical scheme, identical to one obtained in \ac{LBM} methodology,
\begin{linenomath}\begin{equation}
\tilde{\phi}(t+\dt) = \tilde{\phi}(t)+\dt Q(\phi(t),t)  \hspace{2em} \text{where} \hspace{2em} \phi(t) = \tilde\phi(t) +  \frac{\dt}{2}Q(\phi(t),t).
\end{equation}\end{linenomath}

\section{Notes on integration of the \acl{DBE}}\label{app:derivations_DBE}
To ease the reading of the main text, some of the algebraic transformations are listed here.
The substitutions can prove a useful reference for the reader, especially for newcomers, although similar derivations can be also found \ac{LBM} textbooks \cite{Kruger2017}.

Substituting \cref{eq:1defining_tilde_f} into the left-hand side of~\cref{eq:before_defining_tilde_f} 
and~\cref{eq:2defining_tilde_f} into the right-hand side of~\cref{eq:before_defining_tilde_f} gives,
\begin{linenomath}\begin{align}
\lbmh_i(\hat{\bs x},\hat{t}\,)
&=
\left[ 1- \dfrac{1}{2 \tau} \right] \dfrac{ \lbmh_i(\bs x,t) + \dfrac{1}{2\tau}\h^\eq_i(\phi,u\,) + \dfrac{1}{2} q_i(\bs x, t)}{1 + \dfrac{1}{2 \tau}}
+ \dfrac{1}{2\tau}\h^\eq_i(\phi,u\,) + \dfrac{1}{2} q_i(\bs x, t) \nonumber \\[10pt] 
&=
\dfrac{1- \dfrac{1}{2 \tau}}{1 + \dfrac{1}{2 \tau}} \left[ \lbmh_i(\bs x,t) + \dfrac{1}{2\tau}\h^\eq_i(\phi,u\,) + \dfrac{1}{2} q_i(\bs x, t) \right] 
+ \dfrac{1}{2\tau}\h^\eq_i(\phi,u\,) + \dfrac{1}{2} q_i(\bs x, t) \nonumber \\[15pt] 
&= \dfrac{2\tau -1}{2\tau +1}  \left[ \lbmh_i(\bs x,t) + \dfrac{1}{2\tau}\h^\eq_i(\phi,u\,) + \dfrac{1}{2} q_i(\bs x, t) \right] 
+ \dfrac{1}{2\tau}\h^\eq_i(\phi,u\,) + \dfrac{1}{2} q_i(\bs x, t) \nonumber \\[15pt] 
&= \underbrace{\dfrac{2\tau -1}{2\tau +1}}_{F_1} \lbmh_i(\bs x,t) 
+ \underbrace{\dfrac{1}{2\tau} \left[1+\dfrac{2\tau -1}{2\tau +1}\right] }_{F_2} \h^\eq_i(\phi,u\,) 
+  \underbrace{\dfrac{1}{2}\left[1+\dfrac{2\tau -1}{2\tau +1}\right]}_{F_3} q_i(\bs x, t) . \label{eq:waits_for_omega}
\end{align}\end{linenomath}
Next, introducing the relaxation frequency, $\omega=\dfrac{1}{\tau +1/2}$, simplifies the expressions further,
\begin{linenomath}\begin{align}
F_1 &= \dfrac{2\tau -1}{2\tau +1} = \dfrac{2\tau + 1}{2\tau +1} -\dfrac{2}{2\tau +1} = 1 - \dfrac{1}{\tau +1/2} = 1 - \omega. \label{eq:F1} \\[15pt] 
F_2 &= \dfrac{1}{2\tau} \left[1 + \dfrac{2\tau -1}{2\tau +1} \right] = \dfrac{1}{2\tau} \dfrac{4\tau}{2\tau +1} = \dfrac{2}{2\tau +1} = \omega. \label{eq:F2} \\[15pt] 
F_3 &= \dfrac{1}{2} \left[1 + \dfrac{2\tau -1}{2\tau +1} \right] = \dfrac{2\tau }{2\tau +1} = \dfrac{2\tau + 1}{2\tau +1} - \dfrac{2}{2} \dfrac{1}{2\tau +1} = 1 - \dfrac{1}{2} \omega. \label{eq:F3}
\end{align}\end{linenomath}
Substituting Equations (\ref{eq:F1}-\ref{eq:F3}) into~\cref{eq:waits_for_omega}, the desired~\cref{eq:tilde_f_evolution} emerges,
\begin{linenomath}\begin{align}
\lbmh_i(\bs x + \bs e_i,t + 1) = 
\lbmh^\star_i(\bs x,t) 
&= ( 1 - \omega) \lbmh_i(\bs x,t)  + \omega  \h^\eq_i(\phi,u\,)  + \left(1 - \dfrac{\omega}{2} \right)q_i(\bs x, t).
\end{align}\end{linenomath}

\section{Comparison of approaches to the source term integration}\label{app:comparison_of_approaches}
The appropriate integration of the \ac{DBE} with the trapezoidal rule and redefinition of variables allows one to derive a second-order accurate, local, and explicit evolution scheme as presented in \cref{eq:tilde_f_evolution}.
To broaden the discussion of approaches available in the literature, this appendix briefly analyses the spatio-temporal derivatives of the source term that appears in the works of Shi et al.~\cite{Shi2008,Shi2009} and Chai et al.~\cite{Chai2016,Chai2020}.
If the source term is not included in the shift of variables, then the evolution equation reads,
\begin{linenomath}\begin{align}
\overline{h}_i(\bs x + \bs e_i,t + 1) =  \overline{h}_i^{\, \star}(\bs x,t) =  (1 - \omega) \overline{h}_i(\bs x,t) + \omega h^\eq_i(\phi, \bs u) + q_i(\phi, \bs x, t) + \frac{1}{2}\left( \pr{t} + \bs e_i \cdot\nabla \right) q_i(\phi, \bs x, t), \label{eq:scheme2}
\end{align}\end{linenomath}
where contrary to the presented scheme, $\phi = \overline{\phi} = \sum_i \overline{h}_i$ and $\overline{h}$ is a shifted variable, analogous to $\lbmh_i$,  but without the contribution of $q_i$.
 
To understand the relation between~\cref{eq:scheme2}, and the one described in the present study (given by~\cref{eq:tilde_f_evolution}), consider the right hand side of~\cref{eq:DBE_with_source_term}. 
The $I_2$ integral consist of two parts, first related to collision and second to the source term.
In fact, any number of different numerical techniques can be applied to each of them.
For example, instead of applying the trapezoidal rule, 
one can use a first order approximation of the source term,
\begin{linenomath}\begin{align}
q_i\left(\phi(\bs x+s \bs e_i,t+s\right), \bs x+s \bs e_i,t+s) & = q_i\left(\phi(\bs x,t), \bs x,t\right) + s\pr{s}\Big(q_i\left(\phi(\bs x+s \bs e_i,t+s\right), \bs x+s \bs e_i,t+s)\Big) + \mathcal{O}(s^2) \nonumber \\
& = q_i\left(\phi(\bs x,t), \bs x,t\right) + 
s\left( \underbrace{\frac{\partial t}{\partial s}}_{=1} \pr t + \underbrace{\frac{\partial \bs x }{\partial s}}_{= \bs e_i}  \cdot \pr x \right)
q_i\left(\phi(\bs x,t), \bs x,t\right)  + \mathcal{O}(s^2) \nonumber \\
&\simeq q_i\left(\phi(\bs x,t), \bs x,t\right) + 
s\left(  \pr t +  \bs e_i  \cdot \nabla \right)
q_i\left(\phi(\bs x,t), \bs x,t\right).
\end{align}\end{linenomath}
This approximation can be integrated giving,
\begin{linenomath}\begin{align}
\int_0^1 q_i\left(\phi(\bs{x} +s \bs{e_i},t+s\right), \bs{x}+s \bs{e_i},t+s) ds 
& = \bigg[s q_i\left(\phi(\bs{x}, t), \bs{x}, t\right) +  \frac{s^2}{2}\left(\pr t + \bs{e_i} \cdot \nabla \right) q_i\left(\phi(\bs{x}, t), \bs{x}, t\right) + \mathcal{O}(s^3) \; \bigg|_0^1 \nonumber \\
& \simeq q_i(\phi, \bs{x} ,t) + \dfrac{1}{2}\left(\pr t + \bs e_i \cdot \nabla \right) q_i(\phi, \bs{x},t). \label{eq:Taylor_expansion}
\end{align}\end{linenomath} 

In the case of the \textit{bottom-up} approach, the last term in \cref{eq:Taylor_expansion} is usually recognised as an artefact \cite{Seta2013}. 
As pointed out by \citet{Seta2013}, it can be removed by redefinition of variables
or by addition of a correction with regard to the derivative of the source term.
The derivative can be computed using a forward or backward \ac{FD} expression.
As one would expect, if a forward \ac{FD} is used, then the result is equivalent to that obtained through the trapezoidal rule, 
\begin{linenomath}\begin{align}
q_i(\phi, \bs x,t) + \dfrac{1}{2} \left(\pr t + \bs e_i \cdot \nabla \right) q_i(\phi, \bs x,t) 
= q_i(\phi, \bs x,t) + \dfrac{1}{2}\left(q_i(\hat{\phi}, \hat{\bs x}, \hat{t} \,) - q_i(\phi, \bs x,t)\right) 
= \dfrac{1}{2}\left(q_i(\hat{\phi}, \hat{\bs x}, \hat{t} \,) + q_i(\phi, \bs x,t)\right) .
\end{align}\end{linenomath}
The influence of the spatial component of the derivate, $\bs e_i \cdot \nabla$, and various \ac{FD} stencils has been analysed in~\cite{Shi2008}.
As long as the source term does not depend on $\phi$, both forward and backwards \ac{FD} are computationally trivial.
However, if there is a dependence, the backwards \ac{FD} would require additional transfer of data, while the forward \ac{FD}, without the appropriate shift of variables, results in a globally implicit scheme.

\bibliographystyle{model1-num-names}
\bibliography{elsarticle-template-source-term}

\end{document}